\documentclass[10pt, a4paper, twocolumn]{IEEEtran}\newcommand{\pictwidth}{80truemm}

\usepackage{amsmath,epsfig,amssymb,verbatim,amsopn,citesort,color,stfloats}

\DeclareMathOperator{\mvec}{vec}
\newcommand{\ve}[1]{\boldsymbol{#1}}
\newcommand{\E}[1]{E\left\{#1\right\}}
\newcommand{\vA}{\ve{A}} 
\newcommand{\vB}{\ve{B}} 
\newcommand{\vC}{\ve{C}} 
\newcommand{\vD}{\ve{D}} 
 
\newcommand{\vF}{\ve{F}} 
 
\newcommand{\vH}{\ve{H}} \newcommand{\vh}{\ve{h}}
\newcommand{\vI}{\ve{I}} 
\newcommand{\vJ}{\ve{J}}

\newcommand{\vM}{\ve{M}} 
\newcommand{\vN}{\ve{N}} \newcommand{\vn}{\ve{n}}
 
\newcommand{\vP}{\ve{P}} \newcommand{\vp}{\ve{p}}
\newcommand{\vQ}{\ve{Q}} \newcommand{\vq}{\ve{q}}
\newcommand{\vR}{\ve{R}} 
\newcommand{\vS}{\ve{S}} \newcommand{\vs}{\ve{s}}
 
\newcommand{\vU}{\ve{U}} \newcommand{\vu}{\ve{u}}
\newcommand{\vV}{\ve{V}} \newcommand{\vv}{\ve{v}}
\newcommand{\vW}{\ve{W}} \newcommand{\vw}{\ve{w}}
\newcommand{\vX}{\ve{X}} 
\newcommand{\vY}{\ve{Y}} \newcommand{\vy}{\ve{y}}
 
\newcommand{\vDd}{\ve{\Lambda}}

\newcommand{\vXm}{\ve{\mathcal{X}}}
\newcommand{\vSm}{\ve{\mathcal{S}}}
\newcommand{\vDel}{\ve{\Delta}}

\newcommand{\snr}{\overline{\gamma}}

\newcommand{\powd}{\mathcal{P}}

\newcommand{\con}[1]{{#1}^{\ast}}
\newcommand{\mct}[1]{{#1}^{\dagger}}
\newcommand{\mt}[1]{{#1}^{T}}
\newcommand{\RxA}{n_R}
\newcommand{\TxA}{n_T}
\newcommand{\RxM}{N_R}
\newcommand{\TxM}{N_T}

\newcommand{\mcJ}{{\mathcal{J}}}
\newcommand{\aod}{\phi}
\newcommand{\aoa}{\varphi}

\newcommand{\maod}{\aod_0}

\newcommand{\AuthorOne}{Tharaka A. Lamahewa}
\newcommand{\AuthorTwo}{Rodney A. Kennedy}
\newcommand{\AuthorThree}{Thushara D. Abhayapala}
\newcommand{\AuthorFour}{Van K. Nguyen}

\newcommand{\ThankOne}{This work was supported by the Australian Research Council
Discovery Grant DP0343804.}

\newcommand{\ThankTwo}{T. A. Lamahewa, R. A. Kennedy and T. D. Abhayapala are with the
Department of Information Engineering, Research School of
Information Sciences and Engineering, The Australian National
University, Canberra ACT 0200, Australia. email: \{tharaka.lamahewa,
rodney.kennedy, thushara.abhayapala\}@anu.edu.au.}

\newcommand{\ThankThree}{R. A. Kennedy and T. D. Abhayapala are also with
National ICT Australia, Locked Bag 8001, Canberra, ACT 2601,
Australia. National ICT Australia is funded through the Australian
Government's \textit{Backing Australia's Ability} initiative, in
part through the Australian Research Council.}

\newcommand{\ThankFour}{V. K. Nguyen is with the
School of Engineering and Technology, Deakin University, Geelong,
VIC 3217, Australia. email: vknguyen@deakin.edu.au.}

\newcommand{\ThankFive}{Part of this paper has been presented in APCC-2005, Perth, Australia.}

\title{Spatial Precoder Design for Space-Time Coded MIMO Systems: Based on  Fixed Parameters of MIMO Channels}
\author{\authorblockN{\AuthorOne,\:\AuthorTwo,\:\AuthorThree,\:\AuthorFour\thanks{\ThankOne}
\thanks{\ThankTwo\quad\ThankThree\quad\ThankFour}\thanks{\ThankFive}} }

\begin{document}

\maketitle

\begin{abstract}
In this paper, we introduce the novel use of linear spatial
precoding based on fixed and known parameters of multiple-input
multiple-output (MIMO) channels to improve the performance of
space-time coded MIMO systems. We derive linear spatial precoding
schemes for both coherent (channel is known at the receiver) and
non-coherent (channel is un-known at the receiver) space-time coded
MIMO systems. Antenna spacing and antenna placement (geometry) are
considered as fixed parameters of MIMO channels, which are readily
known at the transmitter. These precoding schemes exploit the
antenna placement information at both ends of the MIMO channel to
ameliorate the effect of non-ideal antenna placement on the
performance of space-time coded systems. In these schemes, the
precoder is fixed for given transmit and receive antenna
configurations and transmitter does not require any feedback of
channel state information (partial or full) from the receiver.
Closed form solutions for both precoding schemes are presented for
systems with up to three receiver antennas. A generalized method is
proposed for more than three receiver antennas. We use the coherent
space-time block codes (STBC) and differential space-time block
codes to analyze the performance of proposed precoding schemes.
Simulation results show that at low SNRs, both precoders give
significant performance improvement over a non-precoded system for
small antenna aperture sizes.

\end {abstract}



\section{Introduction}\label{sec:intro}
\PARstart{M}IMO communication systems that use multi-antenna arrays
simultaneously during transmission and reception have generated
significant interest in recent years. Under the assumption of fading
channel coefficients between different antenna elements are
statistically independent and fully known at the receiver (coherent
detection), theoretical work of \cite{telatar-1995} and
\cite{foschini-1998} revealed that the channel capacity of
multiple-antenna array communication systems scales linearly with
the smaller of the number of transmit and receive antennas.
Motivated by these works,
\cite{tarokh_1998_first,Alamouti-1998-block,tarokh_1998_block} have
proposed several modulation and coding schemes, namely space-time
trellis codes and space-time block codes, to exploit the potential
increase in capacity and diversity gains using multi antenna arrays
with coherent detection.

The effectiveness of these coherent space-time coding schemes mainly
relies on the accuracy of the channel estimation at the receiver.
Therefore, differential space-time coding (DSTC) schemes proposed in
\cite{tarokh_diff_stbc,hughes_diff_coding,hochwald_diff_coding} make
an attractive alternative to combat inaccuracy of channel estimation
in coherent space-time coding schemes. With DSTC schemes, channel
state information is not required at either end of the channel.
However, it is well known that DSTC schemes suffer a 3dB performance
loss compared to space-time coding schemes with coherent detection
at the receiver.

For both schemes, code structures are designed assuming that the
channel gains between the transmitter and receiver antennas undergo
uncorrelated independent flat fading. Such an assumption  is valid
only if the scattering environment is isotropic, i.e., scattering is
uniformly distributed over the receiver and transmitter antenna
arrays, and also only if the antennas in an array are well
separated. Recent studies have shown that insufficient antenna
spacing and non-isotropic scattering reduce the performance of
space-time coded communication systems
\cite{Bolcskei_2000,uysal_2001,tharaka_exact_pep_jcn}. This has
motivated the design of linear precoders for space-time coded
multiple antenna systems by exploiting the statistical information
of the MIMO channels
\cite{sampath_precoding,giannakis_precoding,zhao_precoding,gesbert_are_precoding,xiadong_giannakis_diff_precoder,Nguy_icit05}.
In these schemes, the receiver either feeds back the full channel
state information or the correlation coefficients of the channel
(covariance feedback) to the transmitter via a low rate feedback
channel. In order to be cost effective and optimal, these designs
assumed that the channel remains stationary (channel statistics are
invariant) for a large number of symbol periods and the transmitter
is capable of acquiring robust channel state information. However,
when the channel is non-stationary or it is stationary for a small
number of symbol periods, the receiver will have to feedback the
channel information to the transmitter frequently. As a result, the
system becomes costly and the optimum precoder design, based on the
previously possessed information, becomes outdated quickly. In some
circumstances feeding back channel information is not possible.
These facts have motivated us to design a precoding scheme based on
fixed and known parameters of the underlying MIMO channel. Following
list summarizes the original contributions of this paper.

\begin{itemize}
  \item By exploiting the spatial dimension of a MIMO
channel, we design linear spatial precoding schemes to improve the
performance of coherent and differential space-time block coded
systems. These linear spatial precoders are designed based on
previously unutilized fixed and known parameters of MIMO channels,
the antenna spacing and antenna placement details. We use the
spatial channel decomposition given in
\cite{abhayapala_2003_channel} to incorporate the antenna spacing
and antenna placement details into the precoder design.

 \item  Both precoders are fixed for fixed\footnote{antennas are fixed relative to each other} antenna
placement and the transmitter does not require any form of feedback
of channel state information (partial or full) from the receiver.

 \item Since the designs are based on fixed parameters, these spatial
precoders can be used in non-stationary channels as well as
stationary channels.

\item Upper bounds
for pairwise error probability (PEP) of coherent space-time codes
and differential space-time codes are derived for spatially
correlated MIMO channels. To the authors knowledge, the PEP upper
bound of differential space-time codes is a new bound. Utilizing the
MIMO channel decomposition given in \cite{abhayapala_2003_channel},
antenna configuration details and scattering environment parameters
(angular spreads and mean angle of arrival and departure) are
incorporated in to these PEP upper bounds. Assuming an isotropic
scattering environment surrounding the transmitter and receiver
antenna arrays, we minimize the two PEP upper bounds to obtain the
optimum precoders.

 \item We show that our spatial precoding schemes reduce the effect of
non-ideal antenna placement, which is a major contributor to the
spatial correlation, on the MIMO system performance. In these
schemes, the precoder virtually arranges the antennas into an
optimal configuration as such the spatial correlation between all
antenna elements is reduced.

 \item The precoder design is based on the spatial channel model
 proposed in \cite{abhayapala_2003_channel}, but we show that the performance
 of both precoding schemes does not depend on the channel model that used
 to model the underlying MIMO channel. Therefore, our design and simulation
 results provide an independent confirmation of the validity and usefulness of the
 channel model proposed in \cite{abhayapala_2003_channel}.
\end{itemize}

An outline of the paper is as follows. Section
\ref{sec:spatial_channel_model} reviews the spatial channel model
used in our design. In Section \ref{sec:system_model}, the precoded
coherent STBC and differential STBC systems are described along with
detection rules at the receiver. Sections
\ref{sec:problem_setup_coherent_stbc} and
\ref{sec:problem_setup_diff_stbc} present the optimization problem
and the optimal precoder solution for coherent STBC and differential
STBC, respectively. For both precoding schemes, we show that the
optimum linear precoder for a multiple-input single-output (MISO)
fading channel is essentially given by the classical
``water-filling" strategy found in information theory
\cite{telatar-1995}. For a MIMO channel, the linear precoder is
determined by a novel generalized water-filling scheme. Closed form
solutions for both precoding schemes are presented for systems with
up to three receiver antennas. A generalized method is proposed for
more than three receiver antennas. Sections
\ref{sec:simulation_results_coherent_stbc} and
\ref{sec:simulation_results_diff_stbc} present results obtained with
proposed precoding schemes for various spatial scenarios using the
spatial channel model in \cite{abhayapala_2003_channel} as the
underlying MIMO channel. Section
\ref{sec:simulation_results_diff_stbc} also presents results
obtained with proposed precoding scheme for non-isotropic scattering
environments (i.e., limited angular spreads at the transmitter and
receiver antenna arrays). Section
\ref{sec:performance_other_channels} gives the simulation results of
our proposed precoding scheme applied on other statistical channel
models found in the literature. Section \ref{sec:conclusion} present
some concluding remarks and five appendices contain various proofs.

\textbf{\\Notations:} Throughout the paper, the following notations
will be used: Bold lower (upper) letters denote vectors (matrices).
$\mt{[\cdot]}$, $\con{[\cdot]}$ and $\mct{[\cdot]}$ denote the
transpose, complex conjugate and conjugate transpose operations,
respectively. The symbols $\delta(\cdot)$ and $\otimes$ denote the
Dirac delta function and Matrix Kronecker product, respectively. The
notation $\E{\cdot}$ denotes the mathematical expectation,
$\mvec({\vA})$ denotes the vectorization operator which stacks the
columns of $\vA$, $\text{tr}\{\cdot\}$ denotes the matrix trace,
$\lceil{.}\rceil$ denotes the ceiling operator and $\mathbb{S}^1$
denotes the unit circle. The matrix $\vI_n$ is the $n\times{n}$
identity matrix.\\

\section{Spatial Channel Model}\label{sec:spatial_channel_model}
First we review the spatial channel model proposed in
\cite{abhayapala_2003_channel}. Consider a MIMO system consisting of
$\TxA$ transmit antennas located at positions $\vu_t$,
$t=1,2,\cdots,\TxA$ relative to the transmitter array origin, and
$\RxA$ receive antennas located at positions $\vv_r$,
$r=1,2,\cdots,\RxA$ relative to the receiver array origin. $r_T \geq
\max\parallel\!\!\vu_t\!\!\parallel$ and $r_R \geq
\max\parallel\!\!\vv_r\!\!\parallel$ denote the radius of spheres
that contain all the transmitter and receiver antennas,
respectively. We assume that scatterers are distributed in the far
field from the transmitter and receiver antennas and regions
containing the transmit and receive antennas are distinct.

By taking into account physical aspects of scattering, the MIMO
channel matrix $\vH$ can be decomposed into deterministic and random
parts as \cite{abhayapala_2003_channel}
\begin{align}\label{eqn:channel_decompo}
\vH &= \vJ_R\vH_S\mct{\vJ}_T,
\end{align}
where $\vJ_R$ is the deterministic receiver configuration matrix,
\begin{align}\nonumber
\vJ_R &= \left[
  \begin{array}{ccc}
    \mcJ_{-\RxM}(\vv_1) & \cdots & \mcJ_{\RxM}(\vv_1) \\
    \mcJ_{-\RxM}(\vv_2) & \cdots & \mcJ_{\RxM}(\vv_2) \\
    \vdots & \ddots & \vdots \\
    \mcJ_{-\RxM}(\vv_{\RxA}) & \cdots & \mcJ_{\RxM}(\vv_{\RxA}) \\
  \end{array}
\right],
\end{align}
and $\vJ_T$ is the deterministic transmitter configuration matrix,
\begin{align}\nonumber
\vJ_T &= \left[
  \begin{array}{ccc}
    \mcJ_{-\TxM}(\vu_1) & \cdots & \mcJ_{\TxM}(\vu_1) \\
    \mcJ_{-\TxM}(\vu_2) & \cdots & \mcJ_{\TxM}(\vu_2) \\
    \vdots & \ddots & \vdots \\
    \mcJ_{-\TxM}(\vu_{\TxA}) & \cdots & \mcJ_{\TxM}(\vu_{\TxA}) \\
  \end{array}
\right].
\end{align}
$\mcJ_{n}(\vw)$ is the spatial-to-mode function (SMF) which maps the
antenna location $\vw$ to the $n$-th mode of the region. The form
which the SMF takes is related to the shape of the scatterer-free
antenna region. For a circular region in 2-dimensional space, the
SMF is given by a Bessel function of the first kind
\cite{abhayapala_2003_channel} and for a spherical region in
3-dimensional space, the SMF is given by a spherical Bessel function
\cite{abhayapala_2003_3Dchannel}. For a prism-shaped region in
3-dimensional space, the SMF is given by a prolate spheroidal
function \cite{Hanlen_2003}.

Here we consider the situation where the multipath is restricted to
the azimuth plane only (2-D scattering environment), having no field
components arriving at significant elevations. In this case, the SMF
is given by
\begin{align}
\mcJ_{n}(\vw)\,&{\triangleq}\,J_n(k\|\vw\|)e^{{\imath}n(\aod_w-\pi/2)}\nonumber,
\end{align}
where $J_n(\cdot)$ is the Bessel function of integer order $n$,
vector $\vw\:\equiv\:(\|\vw\|,\aod_w)$ in polar coordinates is the
antenna location relative to the origin of the aperture,
$k=2\pi/\lambda$ is the wave number with $\lambda$ being the wave
length and $\imath=\sqrt{-1}$. $\vJ_T$ is $\TxA{\times}(2\TxM+1)$
and $\vJ_R$ is $\RxA{\times}(2\RxM+1)$, where $2\TxM+1$ and
$2\RxM+1$ are the number of effective\footnote{Although there are
infinite number of modes excited by an antenna array, there are only
finite number of modes $(2N+1)$ which have sufficient power to carry
information.} communication modes at the transmit and receive
regions, respectively. Note, $\TxM$ and $\RxM$ are defined by the
size of the regions containing all the transmit and receive
antennas, respectively \cite{H_Jones_2002_b}. In our case,
\begin{align}\nonumber
\TxM&=\left\lceil{\frac{ker_T}{\lambda}}\right\rceil\text{\quad and }\\
\RxM&=\left\lceil{\frac{ker_R}{\lambda}}\right\rceil,\nonumber
\end{align}
where $e\approx{2.7183}$.

Finally, $\vH_S$ is the $(2\RxM+1)\times(2\TxM+1)$ random complex
scattering channel matrix with $(\ell,m)$-th element given by
\begin{align}\label{eqn:H_S}
\left\{\vH_S\right\}_{\ell,m} &=
\iint_{\mathbb{S}^1\times\mathbb{S}^1}\!g(\aod,\aoa)e^{\imath(m-\TxM-1)\phi}
e^{-\imath(\ell-\RxM-1)\psi} \mathrm{d\aod}\mathrm{d\aoa}
\end{align}
representing the complex scattering gain between the $(m-\TxM-1)$-th
mode of the scatter-free transmit region and $(\ell-\RxM-1)$-th mode
of the scatter-free receiver region, where $g(\aod,\aoa)$ is the
effective random complex scattering gain function  for signals with
angle-of-departure $\aod$ from the scatter-free transmitter region
and angle-of-arrival $\aoa$ at the scatter-free receiver region.

The channel matrix decomposition \eqref{eqn:channel_decompo}
separates the channel into three distinct regions of interest: the
scatter-free region around the transmitter antenna array, the
scatter-free region around the receiver antenna array and the
complex random scattering environment which is the complement of the
union of two antenna array regions. Consequently, the MIMO channel
is decomposed into deterministic and random matrices, where
deterministic portions $\vJ_T$ and $\vJ_R$ represent the physical
configuration of the transmitter and the receiver antenna arrays,
respectively, and the random portion represents the complex
scattering environment between the transmitter and the receiver
antenna regions. The reader is referred to
\cite{abhayapala_2003_channel} for more information regarding this
spatial channel model. Note that the precoder design is based on
this channel model, but the performance does not depend on this
model (see Section \ref{sec:performance_other_channels}). That is,
our design and simulations provide an independent confirmation of
the validity and usefulness of this channel model.

\subsection{Spatial Correlation}\label{sec:spatial_correlation}
Suppose transmitter configuration matrix $\vJ_T$ has the singular
value decomposition (svd) $\vJ_T =\vU_T\vDd_T\mct{\vV}_T$ and
receiver configuration matrix $\vJ_R$ has the svd $\vJ_R
=\vU_R\vDd_R\mct{\vV}_R$. Substituting svds of $\vJ_T$ and $\vJ_R$
in \eqref{eqn:channel_decompo} and using the Kronecker product
identity \cite[page 180]{matrix_computations}
$\mvec(\vA\vX\vB)=(\vB^T\otimes\vA)\mvec{(\vX)}$, we obtain
\begin{align}
\vh = \vh_{JS}(\mt{\vU}_R\otimes\mct{\vU}_T),
\end{align}
where $\vh_{JS}=(\mvec{(\vH_{JS}^T)})^T$ with
$\vH_{JS}=\vDd_R\mct{\vV}_R\vH_S\vV_T\mct{\vDd}_T$. Applying the
same Kronecker product identity to $\mvec{(\vH_{JS}^T)}$ yields
$\vh_{JS}=\vh_S[(\con{\vV}_R\mt{\vDd}_R)\otimes(\vV_T\mct{\vDd}_T)]$,
where $\vh_S = (\mvec(\vH_S^T))^T$. Then the covariance matrix
$\vR_{\vH}$ of the MIMO channel $\vH$ is given by
\begin{align}\label{eqn:channel_covar_matrix}
\vR_{\vH}
&\triangleq\E{\mct{\vh}\vh},\nonumber\\
&=(\con{\vU}_R\otimes\vU_T)\vR_{JS}(\mt{\vU}_R\otimes\mct{\vU}_T),
\end{align}
where $\vR_{JS}
=[(\con{\vDd}_R\mt{\vV}_R)\otimes(\vDd_T\mct{\vV}_T)]\vR_S[(\con{\vV}_R\mt{\vDd}_R)\otimes
(\vV_T\mct{\vDd}_T)]$ with $\vR_S=\E{\mct{\vh}_S\vh_S}$.

In this work, our main objective is to design a linear precoder
which compensates for any detrimental effects of non-ideal antenna
placement/configuration on the performance of space-time block
codes. Here we assume that the scattering environment surrounding
the transmitter and the receiver regions is ``rich\footnote{Even
though precoders are derived for rich scattering channels, these
precoders provide significant performance improvements in non rich
scattering channel environments, see Section
\ref{sec:effects_of_non_iso_scatt}.}", i.e., $\vR_S=\vI$. This
assumption yields the simplification
\begin{subequations}\label{eqn:RM_2}
\begin{align}
\vR_{JS} &= [(\con{\vDd}_R\mt{\vV}_R)\otimes(\vDd_T\mct{\vV}_T)][(\con{\vV}_R\mt{\vDd}_R)
\otimes(\vV_T\mct{\vDd}_T)]\label{eqn:RM_2a}\\
         &= (\con{\vDd}_R\mt{\vDd}_R)\otimes(\vDd_T\mct{\vDd}_T),\label{eqn:RM_2b}
\end{align}
\end{subequations}
where \eqref{eqn:RM_2b} follows from \eqref{eqn:RM_2a} by matrix
identity \cite[page 180]{matrix_computations}
$(\vA\otimes\vC)(\vB\otimes\vD)=\vA\vB\otimes\vC\vD$, provided that
the matrix products $\vA\vB$ and $\vC\vD$ exist, and unitary matrix
properties $\mct{\vV}_R\vV_R=\vI$ and $\mct{\vV}_T\vV_T=\vI$.
Substituting \eqref{eqn:RM_2b} into \eqref{eqn:channel_covar_matrix}
gives
\begin{align}\label{eqn:channel_covar_matrix2}
\vR_{\vH} &=
(\con{\vU}_R\otimes\vU_T)(\vR_R\otimes\vR_T)(\mt{\vU}_R\otimes\mct{\vU}_T),
\end{align}
where
\begin{align}\label{eqn:R_T}
\vR_T&=\vDd_T\mct{\vDd}_T
\end{align}
and
\begin{align}\label{eqn:R_R}
\vR_R &=(\vDd_R\mct{\vDd}_R)^T.
\end{align}
Note that both $\vR_R$ and $\vR_T$ are diagonal matrices, where the
diagonal of $\vR_R$ consists of squared singular values of $\vJ_R$
(or eigen-values of $\vJ_R\mct{\vJ}_R$) and  diagonal of $\vR_T$
consists of squared singular values of $\vJ_T$ (or eigen-values of
$\vJ_T\mct{\vJ}_T$).

\section{System Model}\label{sec:system_model}
At time instance $k$, the space time encoder at the transmitter
takes a set of modulated symbols
$\vC(k)=\{c_1(k),c_2(k),\cdots,c_K(k)\}$ and maps them onto an
$\TxA{\times}T$ code word matrix $\vS_{\ell(k)}\in\mathcal{V}$ of
space-time modulated constellation matrices set
$\mathcal{V}=\{\vS_1,\vS_2,\cdots,\vS_L\}$, where $T$ is the code
length, $L=q^{K}$ and $q$ is the size of the constellation from
which $c_n(k)$, $n=1,\cdots,K$ are drawn. By setting
$|c_n(k)|=1/\sqrt{K}$, each code word matrix $\vS_{\ell(k)}$ in
$\mathcal{V}$ will satisfy the property
$\vS_{\ell(k)}\mct{\vS}_{\ell(k)}=\vI_{\TxA}$ for
$\ell(k)\,=\,1,2,\cdots, L$.

In this paper, we mainly focus on the space-time modulated
constellations with the property
\begin{align}\label{eqn:modulation_property_stbc}
(\vS_i-\vS_j)\mct{(\vS_i-\vS_j)} = \beta_{i,j}\vI_{\TxA},
\forall\;{i}\neq{j},
\end{align}
where $\beta_{i,j}$ is a scalar and $\vS_i,\vS_j\in\mathcal{V}$.
Space-time orthogonal designs in \cite{tarokh_1998_block} and some
cyclic and dicyclic space-time modulated constellations in
\cite{hughes_diff_coding} are some examples which satisfy property
\eqref{eqn:modulation_property_stbc} above.

\subsection{Coherent Space-time Block Codes}
Let $\vs_n$ be the $n$-th column of $\vS_i=[\vs_1,
\vs_2,\cdots,\vs_T]\in\mathcal{V}$. At the transmitter, each code
vector $\vs_n$ is multiplied by a $\TxA\times\TxA$ fixed linear
precoder matrix $\vF_c$ before transmitting out from $\TxA$
antennas. Assuming quasi-static fading, the signals received at
$\RxA$ receiver antennas during $T$ symbol periods can be expressed
in matrix form as
\begin{align}\nonumber
\vY(k)&=\sqrt{E_s}\vH\vF_c\vS_{\ell(k)}+\vN(k),
\end{align}
where $E_s$ is the average transmitted signal energy per symbol
period, $\vN(k)$ is the $\RxA{\times}T$ white Gaussian noise matrix
in which elements are zero-mean independent Gaussian distributed
random variables with variance $\sigma_n^2/2$ per dimension and
$\vH$ is the $\RxA\times\TxA$ channel matrix. In this work, we use
the channel decomposition \eqref{eqn:channel_decompo} to represent
the underlying MIMO channel and the elements of scattering channel
matrix $\vH_S$ are modeled as zero-mean complex Gaussian random
variables (Rayleigh fading).

For coherent STBC, we assume that the receiver has perfect channel
state information (CSI) and transmitter has partial CSI (antenna
placement details). At the receiver, the transmitted codeword is
detected by applying the minimum Euclidian distance detection rule:
\begin{align}
\widehat{\vS}_{\ell(k)}&= \arg
\min_{\vS_{\ell(k)}\in\mathcal{V}}\parallel\vy(k)-\sqrt{E_s}\:\widetilde{\vh}
\vSm_{\ell(k)}\parallel^{2}\nonumber\\
&= \arg
\max_{\vS_{\ell(k)}\in\mathcal{V}}\mathrm{Re}\{\widetilde{\vh}\,
\vSm_{\ell(k)}\,\mct{\vy}(k)\},\label{eqn:ML_receiver_coherent}
\end{align}
where $\vy(k) = \mt{(\mvec(\mt{\vY}(k)))}$,
$\vSm_{\ell(k)}=\vI_{\RxA}\otimes\vS_{\ell(k)}$ and $\widetilde{\vh}
= \mt{(\mvec(\mt{\widetilde{\vH}}))}$ with $\widetilde{\vH} =
\vH\vF_c$.

\subsection{Differential Space-time Block Codes}
In this scheme, codeword matrix $\vS_{\ell(k)}$ is differentially
encoded according to the rule
\begin{align}\nonumber
\vX(k) &= \vX(k-1)\vS_{\ell(k)},\text{  for $k=1,2,\cdots$}
\end{align}
with $\vX(0)=\vI_{\TxA}$. Then, each encoded $\vX(k)$ is multiplied
by a $\TxA{\times}\TxA$ fixed linear precoder matrix $\vF_d$ before
transmitting out from $\TxA$ transmit antennas. 
Assuming quasi-static fading, the signals received at $\RxA$
receiver antennas during $\TxA$ symbol periods can be expressed in
matrix form as
\begin{align}\nonumber
\vY(k)&=\sqrt{E_s}\vH\vF_d\vX(k)+\vN(k),
\end{align}
where $\vN(k)$ is the $\RxA{\times}\TxA$ white Gaussian noise matrix
in which elements are zero-mean independent Gaussian distributed
random variables with variance $\sigma_n^2/2$ per complex dimension
and $\vH$ is the $\RxA\times\TxA$ channel matrix, which is modeled
using \eqref{eqn:channel_decompo}.

Assume that the scattering channel matrix $\vH_S$ remains constant
during the reception of two consecutive received signal blocks
$\vY(k-1)$ and $\vY(k)$, which can be expressed in vector (row) form
as
\begin{align}
\vy(k-1) &= \sqrt{E_s}\vh\vXm(k-1) + \vn(k-1),\nonumber\\
\vy(k)&=\sqrt{E_s}\vh\vXm(k) + \vn(k),\nonumber \\
&=\vy(k-1)\vSm_{\ell(k)} + \vw(k),\label{eqn:yk1}
\end{align}
where $\vy(k) = \mt{(\mvec(\mt{\vY(k)}))}$,
$\vXm(k)=\vI_{\RxA}\otimes(\vF_d\vX(k))$, $\vh =
\mt{(\mvec(\mt{\vH}))}$, $\vn(k) = \mt{(\mvec(\mt{\vN(k)}))}$,
$\vSm_{\ell(k)}=\vI_{\RxA}\otimes\vS_{\ell(k)}$ and
$\vw(k)=\vn(k)-\vn(k-1)\vSm_{\ell(k)}$.

For differential STBC, we assume that receiver has no CSI whilst
transmitter has partial CSI (antenna placement details). From
\eqref{eqn:yk1}, the transmitted code word matrix is detected
differentially using the minimum Euclidian distance detection rule:
\begin{align}
\widehat{\vS}_{\ell(k)}&= \arg
\min_{\vS_{\ell(k)}\in\mathcal{V}}\parallel\vy(k)-\vy(k-1)\vSm_{\ell(k)}
\parallel^{2}\nonumber\\
&= \arg
\max_{\vS_{\ell(k)}\in\mathcal{V}}\mathrm{Re}\{\vy(k-1)\vSm_{\ell(k)}\mct{\vy(k)}\}.
\nonumber
\end{align}

\section{Problem Setup: Coherent
STBC}\label{sec:problem_setup_coherent_stbc}
 Assume that perfect
CSI is available at the receiver and also maximum likelihood (ML)
detection is employed at the receiver. Suppose codeword
$\vS_{i}\in\mathcal{V}$ is transmitted, but the ML-decoder
\eqref{eqn:ML_receiver_coherent} chooses codeword
$\vS_j\in\mathcal{V}$, then as shown in the Appendix
\ref{app:pep_bound_derivation_coherent}, the average pairwise error
probability (PEP) is upper bounded by
\begin{align}\label{eqn:corr_up_bound_coherent}
\mathrm{P}(\vS_i\rightarrow\vS_j)\:&{\leq}\frac{1}{\left|\vI_{\TxA\RxA}+\frac{\snr}{4}
\vR_{\vH}[\vI_{\RxA}\otimes\vS_{\Delta}]\right|},
\end{align}
where
$\vS_{\Delta}=\vF_c(\vS_i-\vS_j)\mct{(\vS_i-\vS_j)}\mct{\vF}_c$,
$\vR_{\vH}=\E{\mct{\vh}\vh}$ with row vector
$\vh=(\mvec{(\vH^T)})^T$ and $\snr=E_s/\sigma_n^2$ is the average
symbol energy-to-noise ratio (SNR) at each receiver antenna.
Substituting \eqref{eqn:channel_covar_matrix2} in
\eqref{eqn:corr_up_bound_coherent} and applying the property
\eqref{eqn:modulation_property_stbc} associated with orthogonal
space-time block codes we obtain
\begin{align}\label{eqn:corr_up_bound_coherent_stbc1}
\mathrm{P}(\vS_i\rightarrow\vS_j)\:&{\leq}\:\frac{1}{\left|\vI_{\TxA\RxA}+
\frac{\snr\beta_{k,\ell}}{4}\vR_{RT}[\vI_{\RxA}\otimes{\mct{\vU}_T\vF_c\mct{\vF}_c\vU_T}]\right|},
\end{align}
where we have used the matrix determinant identity
$\left|\vI+\vA\vB\right|=\left|\vI+\vB\vA\right|$ and let
$\vR_{RT}=\vR_R\otimes\vR_T$.\\

\textit{\textbf{Optimization Problem 1}: Find the optimum spatial
precoder $\vF_c$ that minimizes the average PEP upper bound
\eqref{eqn:corr_up_bound_coherent_stbc1} for coherent STBC, subject
to the transmit power constraint
$\mathrm{tr}\{\vF_c\mct{\vF}_c\}=\TxA$, for given transmitter and
receiver antenna configurations in a rich scattering environment.}\\

%

\subsection{Optimum Spatial Precoder: Coherent
STBC}\label{sec:optimum_precoder_coherent_stbc} The linear precoder
$\vF_c$ is designed by minimizing the maximum of all PEP upper
bounds subject to the power constraint
$\text{tr}\{\vF_c\mct{\vF_c}\}=\TxA$. Alternatively, let
\begin{align}\nonumber
\vQ_c
&=\frac{\snr\beta_{k,\ell}}{4}\mct{\vU}_T\vF_c\mct{\vF}_c\vU_T,
\end{align}
then the average PEP bound \eqref{eqn:corr_up_bound_coherent_stbc1}
becomes
\begin{align}\label{eqn:corr_up_bound_coherent_stbc2}
\mathrm{P}(\vS_i\rightarrow\vS_j)\:&{\leq}\:\frac{1}{\left|\vI_{\TxA\RxA}+
[\vR_R\otimes\vR_T][\vI_{\RxA}\otimes\vQ_c]\right|},
\end{align}
and $\vQ_c$ must satisfy the power constraint
$\text{tr}\{\vQ_c\}=\TxA\snr\beta_{k,\ell}/4$. Since $\log(\cdot)$
is a monotonically increasing function, the logarithm of the average
PEP upper bound \eqref{eqn:corr_up_bound_coherent_stbc2} is used as
the objective function to minimize. Note that $\vQ_c$ in
\eqref{eqn:corr_up_bound_coherent_stbc2} is always positive
semi-definite as $\vQ_c=\vB\mct{\vB}$, with
$\vB=\sqrt{(\snr\beta_{k,\ell})/4}\mct{\vU}_T\vF_c$.

Now the optimum $\vQ_c$ is obtained by solving the optimization
problem:
\begin{align}
\min \quad -\log{\left|\vI_{\TxA\RxA}+
(\vR_R\otimes\vR_T)(\vI_{\RxA}\otimes\vQ_c)\right|}\nonumber\\
\text{subject to}\quad \vQ_c\succeq{0},\;
\text{tr}\{\vQ_c\}=\frac{\TxA\snr\beta}{4},\label{eqn:opt_prob_coherent0}
\end{align}
where $\beta=\min_{k\neq\ell}\{\beta_{k,\ell}\}$ over all possible
codewords\footnote{Setting $\beta=\min_{i\neq{j}}\{\beta_{i,j}\}$
will minimize the error probability of the dominant error
event(s).}. By applying Hadamard's inequality on $\left|\vI+
(\vR_R\otimes\vR_T)(\vI\otimes\vQ_c)\right|$ gives that this
determinant is maximized when $(\vR_R\otimes\vR_T)(\vI\otimes\vQ_c)$
is diagonal \cite{telatar-1995}. Therefore $\vQ_c$ must be diagonal
as $\vR_R$ and $\vR_T$ are both diagonal. Since
$(\vR_R\otimes\vR_T)(\vI\otimes\vQ_c)$ is a positive semi-definite
diagonal matrix with non-negative entries on its diagonal,
$\vI+(\vR_R\otimes\vR_T)(\vI\otimes\vQ_c)$ forms a positive definite
matrix. As a result, the objective function of our optimization
problem is convex \cite[page 73]{boyd_convex_opt_2004}. Therefore
the optimization problem \eqref{eqn:opt_prob_coherent0} above is a
convex minimization problem because the objective function and
inequality constraints are convex and equality constraint is affine.

Let $q_i = [\vQ_c]_{i,i}$, $t_i=[\vR_T]_{i,i}$ and
$r_j=[\vR_R]_{j,j}$. Optimization problem
\eqref{eqn:opt_prob_coherent0} then reduces to finding $q_i>0$ such
that
\begin{align}
\min &\quad-\sum_{j=1}^{\RxA}\sum_{i=1}^{\TxA}\log(1+t_iq_ir_j)\nonumber\\
\text{subject to}&\quad \vq\succeq{0},\nonumber\\
&\ve{1}^T\vq=\frac{\TxA\snr\beta}{4}\label{eqn:opt_prob_coherent1}
\end{align}
where $\vq=[q_1,q_2,\cdots,q_{\TxA}]^T$ and $\ve{1}$ denotes the
vector of all ones.

Introducing Lagrange multipliers
$\ve{\lambda}_c\in\mathbb{R}^{\TxA}$ for the inequality constraints
$-\vq\preceq{0}$ and $\upsilon_c\in\mathbb{R}$ for the equality
constraint $\ve{1}^T\vq={\TxA\snr\beta/4}$, we obtain the
Karush-Kuhn-Tucker (K.K.T) conditions
\begin{gather}
\vq\succeq{0},\quad
\ve{\lambda}_c\succeq{0},\quad\ve{1}^T\vq=\frac{\TxA\snr\beta}{4}\nonumber\\
\lambda_iq_i=0, {\quad}i=1,2,\cdots,\TxA\nonumber\\
-\sum_{j=1}^{\RxA}\frac{r_jt_i}{1+r_jt_iq_i}-\lambda_i+\upsilon_c=0,{\quad}i=1,2,
\cdots,\TxA.\label{eqn:kkt_coherent1}
\end{gather}
$\lambda_i$ in \eqref{eqn:kkt_coherent1} can be eliminated since it
acts as a slack variable\footnote{ If $g(x) \leq \upsilon$ is a
constraint inequality, then a variable $\lambda$ with the property
that $g(x) + \lambda = \upsilon$ is called a slack variable
\cite{boyd_convex_opt_2004}. }, giving new K.K.T conditions
\begin{subequations}\label{eqn:kkt_conditions_final_coherent}
\begin{align}
\vq\succeq{0},\quad &\ve{1}^T\vq=\frac{\TxA\snr\beta}{4}\nonumber\\
q_i\left(\upsilon_c-\sum_{j=1}^{\RxA}\frac{r_jt_i}{1+r_jt_iq_i}\right)&=0,
{\quad}i=1,\cdots,\TxA,\label{eqn:kkt_final_a_coherent}\\
\upsilon_c\geq\sum_{j=1}^{\RxA}\frac{r_jt_i}{1+r_jt_iq_i},&{\quad}i=1,\cdots,\TxA.
\label{eqn:kkt_final_b_coherent}
\end{align}
\end{subequations}
For $\RxA=1$, the optimal solution to
\eqref{eqn:kkt_conditions_final_coherent} is given by the classical
``water-filling'' solution found in information theory
\cite{telatar-1995}. The optimal $q_i$ for this case is given in
Section \ref{sec:miso_channel_sol_coherent}. For $\RxA
>1$, the main problem in finding the optimal $q_i$ for given $t_i$
and $r_j, j=1,2,\cdots,\RxA$ is the case that, there are multiple
terms that involve $q_i$ on \eqref{eqn:kkt_final_a_coherent}.
Therefore we can view our optimization problem
\eqref{eqn:opt_prob_coherent1} as a \textit{generalized
water-filling} problem. In fact the optimum $q_i$ for this
optimization problem is given by the solution to a polynomial
obtained from \eqref{eqn:kkt_final_a_coherent}. In Sections
\ref{sec:miso_channel_sol_ntx2_coherent} and
\ref{sec:miso_channel_sol_ntx3_coherent}, we provide closed form
expressions for optimum $q_i$ for $\RxA=2$ and $3$ receiver antennas
and a generalized method which gives optimum $q_i$ for $\RxA>3$ is
discussed in Section \ref{sec:generalized_mimo_sol_coherent}.

As shown above, the optimal $\vQ_c$ is diagonal with
\begin{align}\nonumber
\vQ_c=\mathrm{diag}\{q_1,q_2,\cdots,q_{\TxA}\},
\end{align}
and optimal spatial precoder $\vF_c$ is obtained by forming
\begin{align}\nonumber
\vF_c=\sqrt{\frac{4}{\beta\snr}}\vU_T\vQ^{\frac{1}{2}}_c\mct{\vU}_n,
\end{align}
where $\vU_n$ is any unitary matrix. In this work, we set $\vU_n
=\vI_{\TxA}$.


\subsection{MISO Channel}\label{sec:miso_channel_sol_coherent}
Consider a MISO channel where we have $\TxA$ transmit antennas and a
single receive antenna. The optimization problem involved in this
case is similar to the water-filling problem in information theory,
which has the optimal solution
\begin{align}
q_i &=
\left\{%
\begin{array}{ll}
\frac{1}{\upsilon_c}-\frac{1}{t_i}, & \hbox{$\upsilon_c<t_i$}, \\
0, & \hbox{otherwise,} \\
\end{array}\nonumber%
\right.
\end{align}
where the water-level $1/\upsilon_c$ is chosen to satisfy
\begin{align}
\sum_{i=1}^{\TxA}\max\left(0,\frac{1}{\upsilon_c}-\frac{1}{t_i}\right)=
\frac{\TxA\snr\beta}{4}\nonumber.
\end{align}

\subsection{$\TxA{\times}2$ MIMO Channel}\label{sec:miso_channel_sol_ntx2_coherent}
We now consider the case of $\TxA$ transmit antennas and $\RxA = 2$
receive antennas. As shown in the Appendix \ref{app:two_rx}, the
optimum $q_i$ for this case is
\begin{align}\label{eqn:water_fill_sol_ntx2_coherent}
q_i &=
\left\{%
\begin{array}{ll}
A+\sqrt{K}, & \hbox{$\upsilon_c<t_i(r_1+r_2);$} \\
0, & \hbox{otherwise,} \\
\end{array}%
\right.
\end{align}
where ${\upsilon_c}$ is chosen to satisfy
\begin{align}
\sum_{i=1}^{\TxA}\max\left(0,A+\sqrt{K}\right)=\frac{\TxA\snr\beta}{4},\nonumber
\end{align}
with
\begin{align}
A &=\frac{2r_1r_2t_i^2-{\upsilon_c}t_i(r_1+r_2)}{2{\upsilon_c}r_1r_2t_i^2}\;
\quad\text{and}\nonumber\\
K &=
\frac{\upsilon^2_ct_i^2(r_1-r_2)^2+4r_1^2r_2^2t_i^4}{2{\upsilon_c}r_1r_2t_i^2}
\label{eqn:A_K_coherent}.
\end{align}

\subsection{$\TxA{\times}3$ MIMO Channel}\label{sec:miso_channel_sol_ntx3_coherent}
For the case of $\TxA$ transmit antennas and $\RxA = 3$ receive
antennas, the optimum $q_i$ is given by
\begin{align}\label{eqn:water_fill_sol_ntx3_coherent}
q_i &=
\left\{%
\begin{array}{ll}
-\frac{a_2}{3a_3}+S+T, & \hbox{$\upsilon_c<t_i(r_1+r_2+r_3);$} \\
0, & \hbox{otherwise,} \\
\end{array}%
\right.
\end{align}
where ${\upsilon_c}$ is chosen to satisfy
\begin{align}
\sum_{i=1}^{\TxA}\max\left(0,-\frac{a_2}{3a_3}+S+T\right)=\frac{\TxA\snr\beta}{4},\nonumber
\end{align}
with
\begin{align}
S+T &= \left[R+\sqrt{Q^3+R^2}\right]^{\frac{1}{3}} +
\left[R-\sqrt{Q^3+R^2}\right]^{\frac{1}{3}},\nonumber\\
Q &= \frac{3a_1a_3-a_2^2}{9a_3^2},\quad R =
\frac{9a_1a_2a_3-27a_0a_3^2-2a_2^3}{54a_3^3},\nonumber
\end{align}
$a_3={\upsilon_c}r_1r_2r_3t_i^3$,
$a_2={\upsilon_c}t_i^2(r_1r_2+r_1r_3+r_2r_3)-3r_1r_2r_3t_i^3$,
$a_1={\upsilon_c}t_i(r_1+r_2+r_3)-2t_i^2(r_1r_2+r_1r_3+r_2r_3)$ and
$a_0={\upsilon_c}-t_i(r_1+r_2+r_3)$. A sketch of the proof of
\eqref{eqn:water_fill_sol_ntx3_coherent} is given in the Appendix-\ref{app:three_rx}.\\

\subsection{A Generalized Method}\label{sec:generalized_mimo_sol_coherent}
We now discuss a method which allows to find optimum solution to
\eqref{eqn:opt_prob_coherent1} for a system with $\TxA$ transmit and
$\RxA$ receive antennas. The complementary slackness condition
$\lambda_iq_i=0$ for $i=1,2,\cdots,\TxA$ states that $\lambda_i$ is
zero unless the $i$-th inequality constraint is active at the
optimum. Thus, from \eqref{eqn:kkt_final_a_coherent} we have two
cases: (i) $q_i=0$ for ${\upsilon_c}>t_i\sum_{j=1}^{\RxA}r_j$, (ii)
${\upsilon_c}=\sum_{j=1}^{\RxA}{r_jt_i}/({1+r_jt_iq_i})$ for $q_i>0$
\cite[page 243]{boyd_convex_opt_2004}. For the later case, the
optimum $q_i$ is found by evaluating the roots of $\RxA$-th order
polynomial in $q_i$, where the polynomial is obtained from
${\upsilon_c}=\sum_{j=1}^{\RxA}{r_jt_i}/({1+r_jt_iq_i})$. Since the
objective function of the optimization problem
\eqref{eqn:opt_prob_coherent1} is convex for $\vq>0$, there exist at
least one positive root to the $\RxA$-th order polynomial for
${\upsilon_c}<t_i\sum_{j=1}^{\RxA}r_j$. In the case of multiple
positive roots, the optimum $q_i$ is the one which gives the minimum
to the objective function of \eqref{eqn:opt_prob_coherent1}. In both
cases, ${\upsilon_c}$ is chosen to satisfy the power constraint
$\ve{1}^T\vq={\TxA\snr\beta}/{4}$.

\section{Problem Setup: Differential STBC}\label{sec:problem_setup_diff_stbc}
For the Differential STBC, we again use the average PEP upper bound
to derive the optimum spatial precoder that reduces the effects of
non-ideal antenna placement on the performance of differential STBC.
Below shows the derivation of the average PEP upper bound.

Based on \eqref{eqn:yk1}, the receiver will erroneously select
$\vS_j$ when $\vS_i$ was actually sent as the $k$-th information
matrix if
\begin{align}
\parallel\vy(k)-\vy(k-1)\vSm_j\parallel^2\:&\leq\:\parallel\vy(k)-\vy(k-1)
\vSm_i\parallel^2,\nonumber\\
\vy({k-1})\vD_{i,j}\mct{\vy}({k-1})\:&\leq\:2\mathrm{Re}\{\vw(k)\mct{\vDel}_{i,j}
\mct{\vy}({k-1})\},\label{eqn:ML_diff_1}
\end{align}
where $\vDel_{i,j}=\vSm_j-\vSm_i = \vI_{\RxA}\otimes(\vS_j-\vS_i)$
and $\vD_{i,j}=\vDel_{i,j}\mct{\vDel}_{i,j} =
\vI_{\RxA}\otimes((\vS_i-\vS_j)\mct{(\vS_i-\vS_j)})$. For given
$\vy(k-1)$, the term on the left hand side of \eqref{eqn:ML_diff_1}
is a constant and the term on the right hand side is a Gaussian
random variable. Let
$u=2\mathrm{Re}\{\vw(k)\mct{\vDel}_{i,j}\mct{\vy}(k-1)\}$, then in
the Appendix \ref{app:mean_and_var_of_u} we have shown that $u$ has
the conditional mean
\begin{align}
\bar{m}_{u{\mid}\vy(k-1)} & = \E{u\mid\vy(k-1)}, \nonumber\\
&= 2\mathrm{Re}\left\{\bar{m}_{\vn(k-1)
\mid\vy(k-1)}(\vI-\vSm_{\!i}\mct{\vSm}_{j})\mct{\vy}(k-1)\right\},\nonumber
\end{align}
where
$\bar{m}_{\vn(k-1)\mid\vy(k-1)}=\sigma_n^2\vy(k-1)(\mct{\vXm}(k-1)\vR_{\vH}\vXm(k-1)
+\sigma_n^2\vI_{\TxA\RxA})^{-1}$, and the conditional variance
\begin{align}
\sigma_{u{\mid}\vy(k-1)}^2
&=\E{\parallel{u}-\bar{m}_{u{\mid}\vy(k-1)}\parallel^2\mid\vy(k-1)},\nonumber\\
&=2\vy(k-1)\vDel_{i,j}\nonumber\\
&\times\left(\sigma_n^2\vI+\mct{\vSm}_{i}\Sigma_{\vn(k-1)\mid\vy(k-1)}\vSm_{i}\right)
\mct{\vDel}_{i,j}\mct{\vy}(k-1),\nonumber
\end{align}
where $\Sigma_{\vn(k-1)\mid\vy(k-1)} =
\sigma_n^2(\vI-\sigma_n^2(E_s\mct{\vXm}(k-1)\vR_{\vH}\vXm(k-1)+\sigma_n^2\vI)^{-1})$.
Recall that $\vR_{\vH}$ in $\bar{m}_{\vn(k-1)\mid\vy(k-1)}$ and
$\Sigma_{\vn(k-1)\mid\vy(k-1)}$ is the channel correlation matrix,
defined by \eqref{eqn:channel_covar_matrix} and $\vXm(k)=\vI_{\RxA}\otimes(\vF_d\vX(k))$.\\

Let $d_{i,j}^2 = \vy(k-1)\vD_{i,j}\mct{\vy}(k-1)$. Based on
\eqref{eqn:ML_diff_1}, the PEP condition on received signal
$\vy(k-1)$ is given by
\begin{align}\label{eqn:pep_cond_yk-1}
\mathrm{P}(\vS_i\rightarrow\vS_j\mid\vy(k-1)) &= \mathrm{Pr}(U>d_{i,j}^2),\nonumber\\
&= \int_{d_{i,j}^2}^{\infty}\frac{1}{\sqrt{2\pi}\sigma}
\exp{\left(-\frac{(u-\bar{m})^2}{2\sigma^2}\right)}\mathrm{d}u,\nonumber\\
 &=\mathrm{Q}\left(\frac{d_{i,j}^2-\bar{m}}{\sigma}\right).
\end{align}
In order to obtain unconditional PEP, we need to average
\eqref{eqn:pep_cond_yk-1} with respect to the distribution of
$\vy(k-1)$. Unlike in the coherent STBC case, finding unconditional
PEP from \eqref{eqn:pep_cond_yk-1} poses a much harder problem due
to the non-zero $\bar{m}_{u{\mid}\vy(k-1)}$ and complicated
$\sigma_{u{\mid}\vy(k-1)}^2$. However, at asymptotically high SNRs
(i.e., keeping $E_s$ constant and $\sigma_n^2{\rightarrow}\:0$) the
conditional mean and the conditional variance of $u$ reduce to
$\bar{m}_{u{\mid}\vy(k-1)}\:{=}\:0$ and
$\sigma_{u{\mid}\vy(k-1)}^2\: {=}\:\:4\sigma_n^2d_{i,j}^2$,
respectively. As shown in the Appendix
\ref{app:pep_bound_derivation_noncoherent}, the average PEP can be
upper bounded by
\begin{align}
\mathrm{P}&(\vS_i\rightarrow\vS_j)
\leq&\nonumber\\
&\frac{1}{2}\frac{1}{\left|\vI+\frac{1}{8}
\left(\snr\mct{\vXm(k-1)}\vR_{\vH}\vXm(k-1)+\vI_{\TxA\RxA}\right)\vD_{i,j}\right|},
\label{eqn:upper_bound_diff1}
\end{align}
where $\snr=E_s/\sigma_n^2$ is the average SNR at each receiver
antenna. As for the coherent STBC case, we mainly focus on the
space-time modulated constellations with the property
\eqref{eqn:modulation_property_stbc}. Furthermore, similar to
\cite{hughes_diff_coding,hochwald_diff_coding} we assume that code
length $T=\TxA$. Under this assumption, each code word matrix
$\vS_{i}$ in $\mathcal{V}$ will satisfy the unitary property
$\vS_{i}\mct{\vS}_{i}=\vI$ and $\mct{\vS}_{i}\vS_{i}=\vI$ for
$i\,=\,1,2,\cdots, L$. As a result, $\vX(k)$ will also satisfy the
unitary property $\vX(k)\mct{\vX}(k)=\vI$ and
$\mct{\vX}(k)\vX(k)=\vI$ for $k\,=\,0,1,2,\cdots$. Applying
\eqref{eqn:modulation_property_stbc} on
\eqref{eqn:upper_bound_diff1} and then using the unitary property of
$\vX(k-1)$ and the determinant identity
$\left|\vI+\vA\vB\right|=\left|\vI+\vB\vA\right|$, after straight
forward manipulations, we can simplify the PEP upper bound
\eqref{eqn:upper_bound_diff1} to
\begin{align}
\mathrm{P}&(\vS_i\rightarrow\vS_j)
\leq\frac{1}{2}\frac{\left(\frac{8+\beta{i,j}}{8}\right)^{-\TxA\RxA}}
{\left|\vI+\frac{\beta_{i,j}\snr}{(8+\beta_{i,j})}\vR_{\vH}(\vI_{\RxA}\otimes\vF_d\mct{\vF}_d)
\right|} \label{eqn:upper_bound_diff2}.
\end{align}

As before, we assume that the scattering environment surrounding the
transmitter and receiver antenna arrays is isotropic. Then,
substitution of \eqref{eqn:channel_covar_matrix2} in
\eqref{eqn:upper_bound_diff2} gives
\begin{align}
\mathrm{P}&(\vS_i\rightarrow\vS_j)
\leq&\nonumber\\
&\frac{1}{2}\frac{\left(\frac{8+\beta{i,j}}{8}\right)^{-\TxA\RxA}}
{\left|\vI+\frac{\beta_{i,j}\snr}{(8+\beta_{i,j})}(\vR_R\otimes\vR_T)
(\vI_{\RxA}\otimes\mct{\vU}_T\vF_d\mct{\vF}_d\vU_T) \right|}
\label{eqn:upper_bound_diff3},
\end{align}
where $\vR_T$ and $\vR_R$  are defined by \eqref{eqn:R_T} and
\eqref{eqn:R_R}, respectively. The optimization problem for
differential STBC case can now be stated as follows:\\

\textit{\textbf{Optimization Problem 2}: Find the optimum spatial
precoder $\vF_d$ that minimizes the average PEP upper bound
\eqref{eqn:upper_bound_diff3} for differential STBC, subject to the
transmit power constraint $\mathrm{tr}\{\vF_d\mct{\vF}_d\}=\TxA$,
for given transmitter and receiver antenna
configurations in a rich scattering environment.}\\

\subsection{Optimum Spatial Precoder: Differential
STBC}\label{sec:optimum_precoder_diff_stbc} Similar to the coherent
STBC case, the optimum spatial precoder $\vF_d$ for differential
STBC is obtained by minimizing the maximum of all PEP upper bounds
subject to the power constraint
$\text{tr}\{\vF_d\mct{\vF}_d\}=\TxA$. Let
\begin{align}\nonumber
\vP_d =
\frac{\beta_{i,j}\snr}{{(8+\beta_{i,j})}}\mct{\vU}_T\vF_d\mct{\vF}_d\vU_T.
\end{align}
The optimum $\vP_d$ (hence the optimum $\vF_d$) is then obtained by
solving the optimization problem
\begin{align}
\min \quad -&\log{\left|\vI+(\vR_R\otimes\vR_T)
(\vI_{\RxA}\otimes\vP_d)\right|}\nonumber\\
\text{subject to}&\quad \vP_d\succeq{0},\;
\text{tr}\{\vP_d\}=\frac{\beta_{i,j}\snr\TxA}{(8+\beta_{i,j})}.\nonumber
\end{align}

The above optimization problem is identical to the optimization
problem derived for coherent STBC, except a different scalar for the
equality constraint. Therefore, following Section
\ref{sec:optimum_precoder_coherent_stbc}, here we present the final
optimization problem and solutions to it without detail derivations.

Following Section \ref{sec:optimum_precoder_coherent_stbc}, we can
show that the optimum $\vP_d$ is diagonal and diagonal entries of
$\vP_d$ are found by solving the optimization problem
\begin{align}
\min &\quad-\sum_{j=1}^{\RxA}\sum_{i=1}^{\TxA}\log(1+t_ip_ir_j)\nonumber\\
\text{subject to}&\quad \vp\succeq{0},\nonumber\\
&\ve{1}^T\vp=\frac{\beta\snr\TxA}{(8+\beta)}\label{eqn:opt_prob_diff2}
\end{align}
where $\beta=\min_{i\neq{j}}\{\beta_{i,j}\}$ over all possible
codewords, $p_i = [\vP_d]_{i,i}$, $t_i=[\vR_T]_{i,i}$
$r_j=[\vR_R]_{j,j}$ and $\vp=[p_1,p_2,\cdots,p_{\TxA}]^T$. The
linear spatial precoder $\vF_d$ is obtained by forming
\begin{align}\nonumber
\vF_d &=
\sqrt{\frac{8+\beta}{\beta\snr}}\vU_T\vP_d^{\frac{1}{2}}\mct{\vU}_n,
\end{align}
where $P_d=\mathrm{diag}\{p_1,p_2,\cdots,p_{\TxA}\}$ and $\vU_n$ is
any unitary matrix. Similar to coherent STBC case, when $\RxA=1$,
the optimum power loading strategy is identical to the
``water-filling" in information theory. When $\RxA>1$, a
\textit{generalized water-filling} strategy gives the optimum
$\vP_d$. Following Sections give the optimum $p_i$ for
\eqref{eqn:opt_prob_diff2} for $\RxA\:=\:1, 2, 3$ receive antennas.
For other cases, the the generalized method discussed in Section
\ref{sec:generalized_mimo_sol_coherent} can be directly applied to
obtain the optimum $p_i$ for \eqref{eqn:opt_prob_diff2}.

\subsection{MISO Channel}\label{sec:miso_channel_sol_diff}
The optimization problem involved in this case is similar to the
water-filling problem in information theory, which has the optimal
solution
\begin{align}\label{eqn:water_fill_sol_ntx1_diff}
p_i &=
\left\{%
\begin{array}{ll}
\frac{1}{\upsilon_d}-\frac{1}{t_i}, & \hbox{$\upsilon_d<t_i$}, \\
0, & \hbox{otherwise,} \\
\end{array}
\right.
\end{align}
where the water-level $1/\upsilon_d$ is chosen to satisfy
\begin{align}
\sum_{i=1}^{\TxA}\max\left(0,\frac{1}{\upsilon_d}-\frac{1}{t_i}\right)=
\frac{\snr\beta\TxA}{8+\beta}\nonumber.
\end{align}

\subsection{$\TxA{\times}2$ MIMO Channel}\label{sec:mimo_channel_sol_ntx2_diff}
The optimum $p_i$ for this case is
\begin{align}\nonumber
p_i &=
\left\{%
\begin{array}{ll}
A+\sqrt{K}, & \hbox{$\upsilon_d<t_i(r_1+r_2);$} \\
0, & \hbox{otherwise,} \\
\end{array}%
\right.
\end{align}
where ${\upsilon}$ is chosen to satisfy
\begin{align}\nonumber
\sum_{i=1}^{\TxA}\max\left(0,A+\sqrt{K}\right)=\frac{\snr\beta\TxA}{8+\beta}
\end{align}
with
\begin{align}\nonumber
A=\frac{2r_1r_2t_i^2-{\upsilon_d}t_i(r_1+r_2)}{2{\upsilon_d}r_1r_2t_i^2},
\end{align}
and
\begin{align}\nonumber
K =
\frac{\upsilon_d^2t_i^2(r_1-r_2)^2+4r_1^2r_2^2t_i^4}{2{\upsilon_d}r_1r_2t_i^2}.
\end{align}

\subsection{$\TxA{\times}3$ MIMO Channel}\label{sec:miso_channel_sol_ntx3_diff}
For the case of $\TxA$ transmit antennas and $\RxA = 3$ receive
antennas, the optimum $p_i$ is given by
\begin{align}\nonumber
p_i &=
\left\{%
\begin{array}{ll}
-\frac{z_2}{3z_3}+Z, & \hbox{$\upsilon_d<t_i(r_1+r_2+r_3);$} \\
0, & \hbox{otherwise,} \\
\end{array}%
\right.
\end{align}
where ${\upsilon}_d$ is chosen to satisfy
\begin{align}\nonumber
\sum_{i=1}^{\TxA}\max\left(0,-\frac{z_2}{3z_3}+Z\right)=\frac{\snr\beta\TxA}{8+\beta},
\end{align}
with
\begin{align}
Z &= \left[Z_2+\sqrt{Z_1^3+Z_2^2}\right]^{\frac{1}{3}} +
\left[Z_2-\sqrt{Z_1^3+Z_2^2}\right]^{\frac{1}{3}},\nonumber\\
Z_1 &= \frac{3z_1z_3-z_2^2}{9z_3^2},\quad Z_2 =
\frac{9z_1z_2z_3-27z_0z_3^2-2z_2^3}{54z_3^3},\nonumber
\end{align}
$z_3={\upsilon_d}r_1r_2r_3t_i^3$,
$z_2={\upsilon_d}t_i^2(r_1r_2+r_1r_3+r_2r_3)-3r_1r_2r_3t_i^3$,
$z_1={\upsilon_d}t_i(r_1+r_2+r_3)-2t_i^2(r_1r_2+r_1r_3+r_2r_3)$ and
$z_0={\upsilon_d}-t_i(r_1+r_2+r_3)$.

\subsection{Spatially Uncorrelated Receive
Antennas}\label{sec:spatially_uncor_rx_antennas} If $\RxA$ receive
antennas are placed ideally within the receiver region such that the
spatial correlation between antenna elements is zero (i.e.,
$\mct{\vJ}_R\vJ_R=\vI$), then the objective function in
\eqref{eqn:opt_prob_diff2} reduces to a single summation and the
optimum $p_i$ is given by the water-filling solution
\eqref{eqn:water_fill_sol_ntx1_diff} obtained for the MISO channel.
This is not to say that such an ideal placement is possible even
approximately. A similar result holds for the coherent STBC case.\\

\section{Simulation Results: Coherent
STBC}\label{sec:simulation_results_coherent_stbc} In this section,
we will illustrate the performance improvements obtained from
coherent STBC when the spatial precoder $\vF_c$ derived in
Section-\ref{sec:optimum_precoder_coherent_stbc} is used. In
particular, the performance is evaluated for small antenna
separations and different antenna geometries at the transmitter and
receiver antenna arrays, assuming an isotropic scattering
environment (independent and identically distributed entries in
scattering channel matrix $\vH_S$). In our simulations we use the
rate-1 space-time modulated constellation constructed in
\cite{tarokh_1998_block} from orthogonal designs for two transmit
antennas. Also use the rate $3/4$ STBC code for $\TxA=3, 4$ transmit
antennas given in \cite{tarokh_1998_block}. Modulated symbols $c(k)$
are drawn from the normalized QPSK alphabet
$\{\pm1/\sqrt{2}\pm{i}/\sqrt{2}\}$.

\subsection{MISO Channels}\label{sec:sim_results_MISO_coherent}
First we illustrate the water-filling concept for $\TxA=2,3$ and $4$
transmit antennas, where the transmit antennas are placed in uniform
circular array (UCA) and uniform linear array (ULA)
configurations\footnote{This precoder can be applied to any
arbitrary antenna configuration.} with $0.2\lambda$ minimum
separation between two adjacent antenna elements. For each transmit
antenna configuration we consider,
Table-\ref{Tab:ant_conf_details_coherent} lists the radius of the
transmit aperture, number of effective communication
modes\footnote{The set of modes form a basis of functions for
representing a multipath wave field.}\cite{abhayapala_2003_channel}
in the transmit region and the rank of the transmit side spatial
correlation matrix $\vJ_T\mct{\vJ}_T$. Note that, in all spatial
scenarios, we ensure that $\vJ_T\mct{\vJ}_T$ is full rank in order
that the average
PEP upper bound \eqref{eqn:corr_up_bound_coherent} to hold.\\

\begin{table}
\caption{Transmit antenna configuration details corresponding to
water-filling scenarios considered in Fig.
\ref{fig:water_filling_coherent}.}
\begin{center}
\begin{tabular}{|c|c|c|c|}
  \hline
  Antenna       & Tx aperture       & Num. of & $\mathrm{rank}(\vJ_T\mct{\vJ}_T)$\\
  Configuration & radius            & modes   & \\
  \hline
                &                   &         &  \\
  2-Tx          & 0.1$\lambda$      & 3       & 2\\
  3-Tx UCA      & 0.115$\lambda$    & 3       & 3\\
  3-Tx ULA      & 0.2$\lambda$      & 5       & 3\\
  4-Tx UCA      & $0.142\lambda$ & 5       & 4\\
  4-Tx ULA      & 0.3$\lambda$      & 7       & 4\\
  \hline
\end{tabular}
\end{center}
\label{Tab:ant_conf_details_coherent}
\end{table}

\begin{figure}[h]
\centering
\includegraphics[width=\pictwidth]{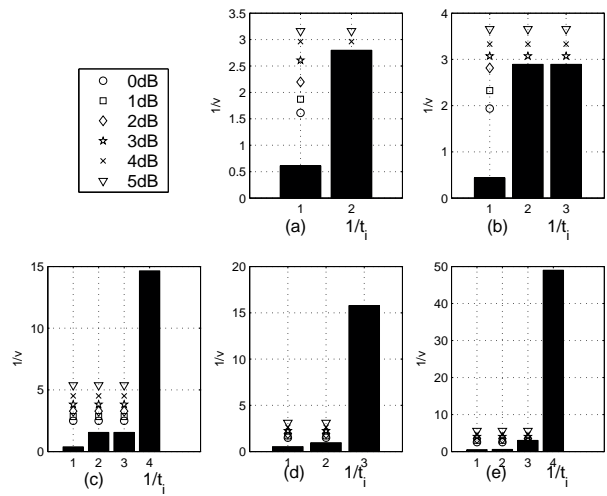}
\caption{Water level ($1/\upsilon_c$) for various SNRs for a MISO
system. (a) $\TxA = 2$, (b) $\TxA = 3$ - UCA, (c) $\TxA = 4$ - UCA,
(d) $\TxA = 3$ - ULA and (e) $\TxA = 4$ - ULA for $0.2\lambda$
minimum separation between two adjacent transmit antennas.}
\label{fig:water_filling_coherent}
\end{figure}

\begin{figure}[h]
\centering
\includegraphics[width=\pictwidth]{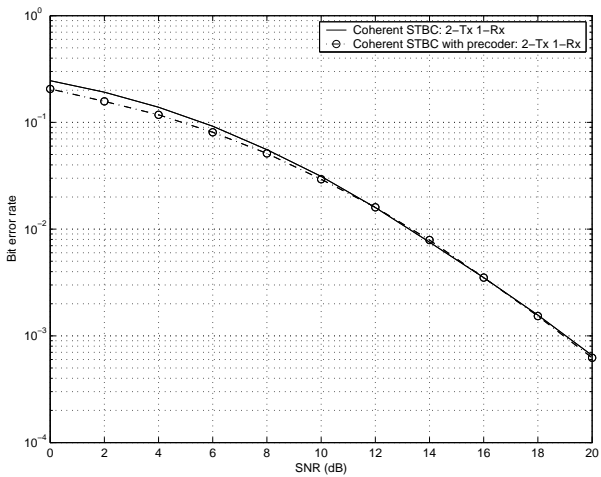}
\caption{Performance of spatial precoder with two transmit and one
receive antennas for $0.2\lambda$ separation between two transmit
antenna elements: rate-$1$ coherent STBC.}
\label{fig:2tx_1rx_coherent}
\end{figure}

Fig. \ref{fig:water_filling_coherent} shows the water levels for
various SNRs. For a given SNR, the optimal power value $q_i$ is the
difference between water-level $1/\upsilon_c$ and base level
$1/t_i$, whenever the difference is positive; it is zero otherwise.
Note that, with this spatial precoder, the diversity order of the
system is determined by the number of non-zero $q_i$'s. It is
observed that at low SNRs, only one $q_i$ is non-zero for $\TxA=2$
and 3-UCA cases. In these cases, all the available power is assigned
to the highest eigen-mode of $\vJ_T\mct{\vJ}_T$ (or to the single
dominant eigen-channel of $\vH$) and the system is operating in
eigen-beamforming mode. With other cases, Fig.
\ref{fig:water_filling_coherent}(c), (d) and (e), systems are
operating in between eigen-beam forming and full diversity for small
SNRs as well as moderate SNRs. In these cases, the spatial precoder
assigns more power to the higher eigen-modes of $\vJ_T\mct{\vJ}_T$
(or to dominant eigen-channels of $\vH$) and less power to the
weaker eigen-modes (or to less dominant eigen-channels of $\vH$).

Fig. \ref{fig:2tx_1rx_coherent} illustrates the BER performance of
the rate 1 STBC with and without spatial precoder for $\TxA=2$. It
can be observed that at very low SNRs, we obtain a pre-coding gain
of about 1.5dB. In fact, at very low SNRs, the optimum scheme is
equivalent to eigen-beam forming. However, as the SNR increases, the
precoder becomes redundant and the optimum scheme approaches STBC,
where it operates in full diversity. This corroborates the claim
that the $2\times{1}$ STBC has good resistance against the spatially
correlated fading at high SNRs as shown in
\cite{tharaka_stbc_fading_2004}.

\begin{figure}[h]
\centering
\includegraphics[width=\pictwidth]{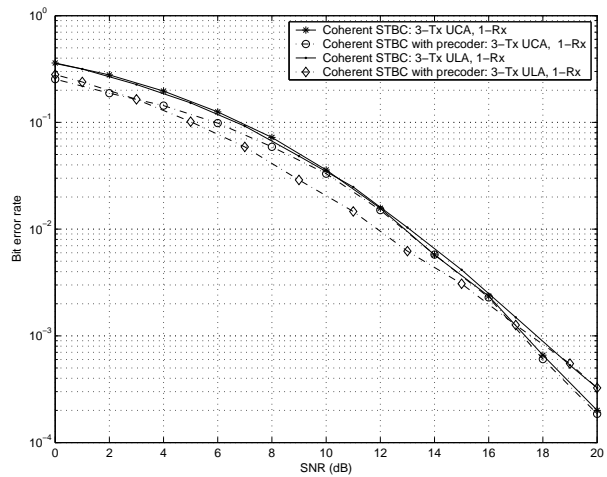}
\caption{Performance of spatial precoder with three transmit and one
receive antennas for $0.2\lambda$ minimum separation between two
adjacent transmit antennas for UCA and ULA antenna configurations:
rate-$3/4$ coherent STBC.} \label{fig:3tx_1rx_coherent}
\end{figure}

\begin{figure}[h]
\centering
\includegraphics[width=\pictwidth]{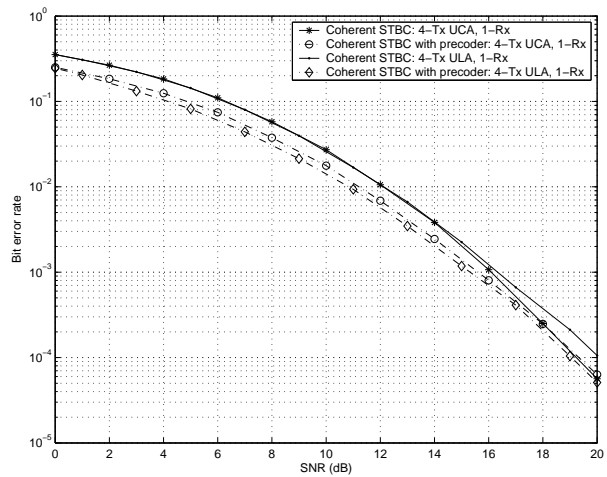}
\caption{Performance of spatial precoder with four transmit and one
receive antennas for $0.2\lambda$ minimum separation between two
adjacent transmit antennas for UCA and ULA antenna configurations:
rate-$3/4$ coherent STBC.} \label{fig:4tx_1rx_coherent}
\end{figure}

BER performance results for 3-Tx UCA, ULA and 4-Tx UCA, ULA antenna
configurations are shown in Fig. \ref{fig:3tx_1rx_coherent} and
\ref{fig:4tx_1rx_coherent}, respectively for rate $3/4$ STBCs. For
3-Tx UCA, the results obtained are similar to the results of $\TxA =
2$ case above. In this case, at low SNRs, the system operates in
eigen beam-forming mode and at high SNRs, it is operating in full
diversity mode as shown in Fig. \ref{fig:water_filling_coherent}(b).
For the other three cases, it is observed that the optimum scheme
provides a clear performance advantage over the STBC only system for
all SNRs concerned. For example, at 0.01 bit-error-rate, we obtain a
precoding gain of about 1dB. However, these systems operate in
between eigen beam-forming and full diversity as the precoder
assigns zero powers to some of the transmit diversity branches of
the channel. As before, at higher SNRs, the system operates in full
diversity and the optimum scheme approaches STBC.

In all cases, at high SNRs we observed that ULA antenna
configuration provides better performance than UCA antenna
configuration when the spatial precoder is used. This is because,
the number of effective communication modes in the transmit region
is higher for the ULA case (large aperture radius of ULA, c.f. Table
\ref{Tab:ant_conf_details_coherent}) than the UCA case and the
spatial precoder efficiently activates the modes in the transmit
region of ULA. This observation suggests that our precoding scheme
gives scope for improvement of ULA performance at high SNR.

\subsection{MIMO Channels}\label{sec:sim_results_MIMO_coherent}
\begin{figure}[h]
\centering
\includegraphics[width=\pictwidth]{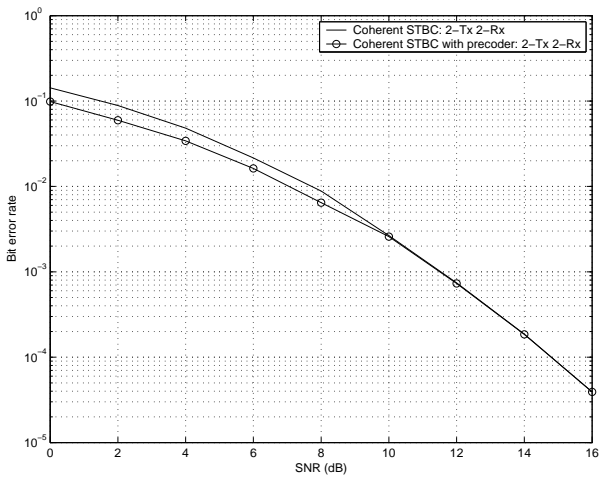}
\caption{Performance of spatial precoder with two transmit and one
receive antennas: receive antenna separation $\lambda$ and minimum
transmit antenna separation $0.2\lambda$ for UCA and ULA antenna
configurations: rate-$3/4$ coherent STBC.}
\label{fig:2tx_2rx_coherent}
\end{figure}

\begin{figure}[h]
\centering
\includegraphics[width=\pictwidth]{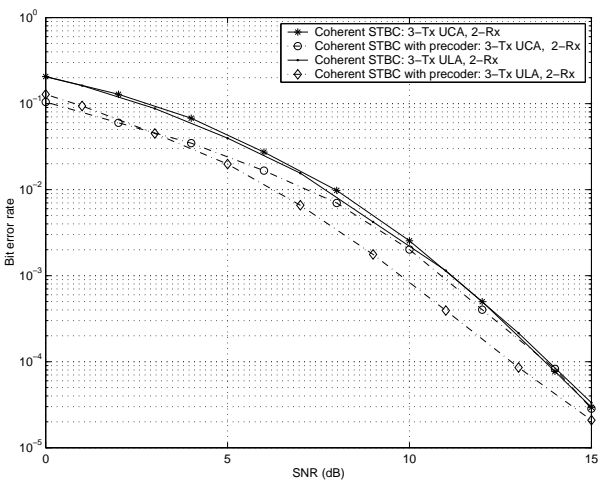}
\caption{Performance of spatial precoder with three transmit and two
receive antennas: receive antenna separation $\lambda$ and minimum
transmit antenna separation $0.2\lambda$ for UCA and ULA antenna
configurations: rate-$3/4$ coherent STBC.}
\label{fig:3tx_2rx_coherent}
\end{figure}

We now examine the performance of the spatial precoder for multiple
transmit and multiple receive antennas. For example, we consider
$\TxA=2,3$ transmit antennas and $\RxA=2$ receive antennas. In all
cases, two receiver antennas are placed $\lambda$ apart, which gives
negligible effects on the performance due to antenna spacing. As
before, the minimum separation between two adjacent transmit
antennas is set to $0.2\lambda$. Note that this situation reasonably
models the uplink of a mobile communication system. For each case,
the optimum $q_i$ is calculated using
\eqref{eqn:water_fill_sol_ntx2_coherent}. Fig.
\ref{fig:2tx_2rx_coherent} illustrates the BER performance results
for 2-transmit, 2-receive antennas for rate $1$ STBC and Fig.
\ref{fig:3tx_2rx_coherent} illustrates the BER performance results
for 3-transmit, 2-receive antennas for rate $3/4$ STBC. Performance
results obtained here are similar to that of MISO cases above.

\section{Simulation Results: Differential
STBC}\label{sec:simulation_results_diff_stbc} We now demonstrate the
performance improvements obtained from differential space time block
coded systems when the spatial precoder derived in Section
\ref{sec:optimum_precoder_diff_stbc} is applied. As before, the
performance of differential space-time coded systems is investigated
for small antenna separations and different antenna geometries
assuming a rich scattering environment surrounding the transmit and
receive antenna arrays (i.e., i.i.d entries in $\vH_S$). We use the
rate-$1$ space-time modulated constellations constructed in
\cite{tarokh_1998_block} from orthogonal designs for two and four
transmit antennas. Normalized QPSK alphabet
$\{\pm1/\sqrt{2}\pm{i}/\sqrt{2}\}$ and normalized BPSK alphabet
$\{\pm1/\sqrt{2}\}$ are used with two and four transmit antenna
STBC, respectively.

\subsection{MISO Channel}\label{sec:sim_results_MISO_diff}
\begin{figure}
\centering
  \includegraphics[width=\pictwidth]{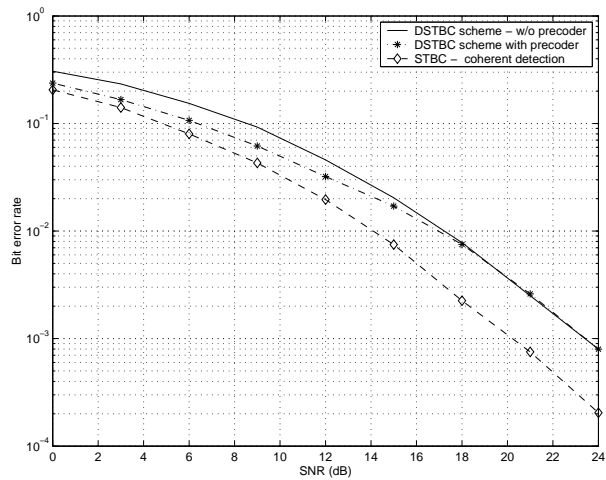}
  \caption{Performance of spatial precoder with two transmit and one receive antennas for
  $0.1\lambda$ separation between two transmit antennas: rate-$1$
  differential STBC.}\label{fig:2tx_1rx_diff}
\end{figure}

Fig. \ref{fig:2tx_1rx_diff} illustrates the BER performance of the
differential STBC with and without spatial precoder when $\TxA= 2$.
Also shown for comparison is the BER performance of the STBC when
coherent detection is employed at the receiver. In all cases, two
transmit antennas are placed $0.1\lambda$ distance apart. It can be
seen that at the BER of $0.05$, the performance of the precoded
system is $1.25$dB better than that of the non-precoded differential
orthogonal space-time coded system and $1.75$dB away from the
coherent detection case. However at high SNRs, the precoder becomes
redundant and the optimum scheme approaches differential STBC.

\begin{figure}
\centering
  \includegraphics[width=\pictwidth]{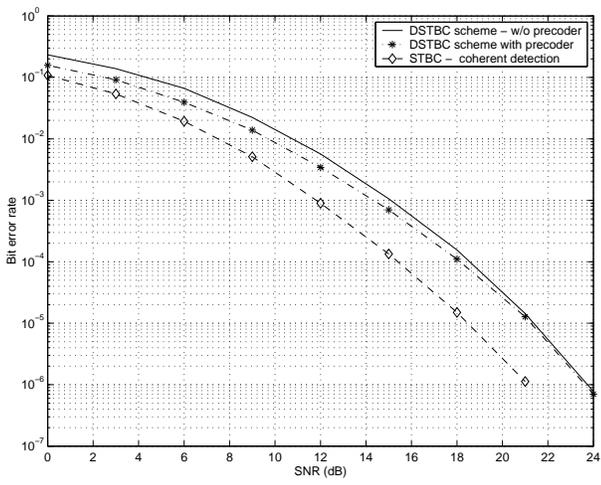}
  \caption{Performance of spatial precoder with four transmit and one receive antennas for
  $0.2\lambda$ minimum separation between two adjacent transmit antennas; UCA transmit antenna
  configuration: rate-$1$ differential STBC.}\label{fig:4tx_uca_1rx_diff}
\end{figure}

\begin{figure}
\centering
  \includegraphics[width=\pictwidth]{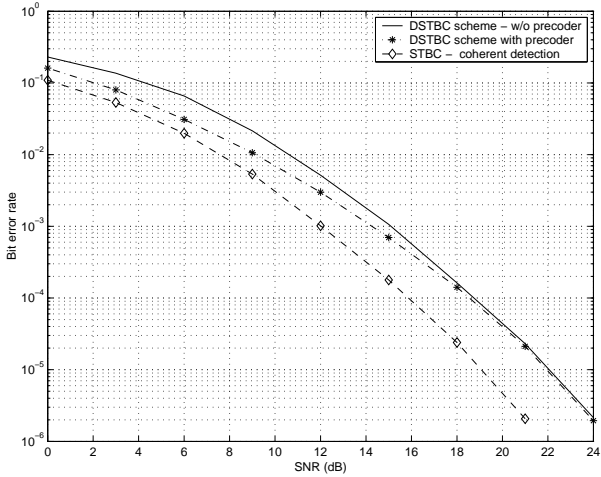}
  \caption{Performance of spatial precoder with four transmit and one receive antennas for
  $0.2\lambda$ minimum separation between two adjacent transmit antennas; ULA transmit antenna
  configuration: rate-$1$ differential STBC.}\label{fig:4tx_ula_1rx_diff}
\end{figure}

BER performance results for 4-Tx UCA and 4-Tx ULA antenna
configurations are shown in Fig. \ref{fig:4tx_uca_1rx_diff} and
\ref{fig:4tx_ula_1rx_diff}, respectively. For both antenna
configurations, the minimum separation between two adjacent antenna
elements is set to $0.2\lambda$, corresponding to aperture radii
$0.142\lambda$ and $0.3\lambda$ for UCA and ULA antenna
configurations, respectively. Simulation results show that the BER
performance of the optimum scheme is better than that of the
differential STBC system for both antenna configurations. For
example, at $10^{-2}$ BER, we obtain precoding gains of about $1$dB
and $1.5$dB with UCA and ULA antenna configurations, respectively.
In comparison with the coherent detection at the receiver, BER
performance of the optimum scheme is $2$dB and $1.5$dB away for UCA
and ULA antenna configurations, respectively.

\subsection{MIMO Channel}\label{sec:sim_results_MIMO_diff}
\begin{figure}
\centering
  \includegraphics[width=\pictwidth]{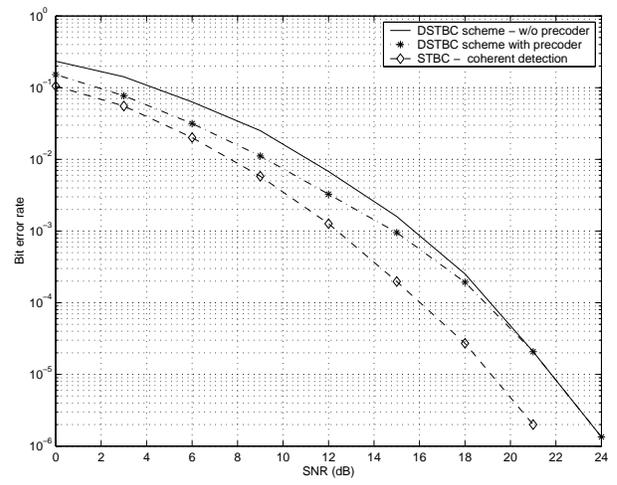}
  \caption{Performance of spatial precoder with two transmit and two receive
  antennas. Transmit antenna separation $0.1\lambda$ and receive antenna separation
  $\lambda$: rate-$1$ differential STBC.
  }\label{fig:2tx_2rx_diff}
\end{figure}

We now examine the performance of the proposed optimum scheme for
multiple transmit and multiple receive antennas. As an example, we
consider a MIMO system consisting of $\TxA=2$ transmit antennas and
$\RxA=2$ receive antennas. The two receiver antennas are placed
$\lambda$ apart, which gives minimum effect on the performance due
to antenna spacing at the receiver antenna array, and the two
transmit antennas are placed $0.1\lambda$ distance apart. Note that
this situation reasonably models the uplink of a mobile
communication system. Fig. \ref{fig:2tx_2rx_diff} shows the
performance of the optimum scheme with two transmit and two receive
antennas. Performance results obtained here are similar to that of
MISO cases considered above.

Note the objective function of D-STBC optimization problem is
derived for high SNR. However, from our simulation results, we
observed that proposed precoding scheme also gives good performance
at low SNRs.

\subsection{Effects of Non-isotropic Scattering}\label{sec:effects_of_non_iso_scatt}
In practise, wireless channels experience non-isotropic scattering
(limited angular spread about a mean angle of departure/arrival)
both at the transmitter and the receiver antenna arrays. We now
investigate the effects of non-isotropic scattering on the BER
performance of differential STBC when the spatial precoding scheme
derived in Section \ref{sec:optimum_precoder_diff_stbc} is used.

First we derive expressions for correlation between different
communication modes at the transmitter and receiver apertures. Using
\eqref{eqn:H_S}, we define the modal correlation between complex
scattering gains as

\begin{align}\nonumber
\gamma_{m,m^{\prime}}^{\ell,\ell^{\prime}}&\triangleq
\E{\{\vH_S\}_{\ell,m}\con{\{\vH_S\}}_{\ell^{\prime},m^{\prime}}}.
\end{align}
Assume that the scattering from one direction is independent of that
from another direction for both the receiver and the transmitter
apertures. Then the second order statistics of the scattering gain
function $g(\aod,\aoa)$ can be defined as
\begin{align}\nonumber
\E{g(\aod,\aoa)\con{g}({\aod}^{\prime},{\aoa}^{\prime})} &\triangleq
G(\aod,\aoa)\delta(\aod-{\aod}^{\prime})\delta(\aoa-{\aoa}^{\prime}),
\end{align}
where $G(\aod,\aoa)=\E{|g(\aod,\aoa)|^2}$ with normalization
${\iint}G(\aod,\aoa)d\aoa{d\aod} = 1$. With the above assumption,
the modal correlation coefficient,
$\gamma_{m,m^{\prime}}^{\ell,\ell^{\prime}}$ can be simplified to
\begin{align}\nonumber
\gamma_{m,m^{\prime}}^{\ell,\ell^{\prime}}&=\iint_{\mathbb{S}^1\times\mathbb{S}^1}G(\aod,\aoa)e^{-i(\ell-\ell^{\prime})
\aoa}e^{i(m-m^{\prime})\aod} d\aoa{d\aod}.
\end{align}
Then the correlation between $\ell$-th and $\ell^{\prime}$-th modes
at the receiver region due to the $m$-th mode at the transmitter
region is given by
\begin{align}\label{eqn:rx_modal_corr_coef}
\gamma^{\ell,\ell^{\prime}}&={\int_{\mathbb{S}^1}}{\powd}_{Rx}(\aoa)
e^{-i(\ell-\ell^{\prime})\aoa}d\aoa,\quad\forall{\,\,m},
\end{align}
where ${\powd}_{Rx}(\aoa)={\int}G(\aod,\aoa){d\aod}$ is the
normalized azimuth power distribution (APD) of the scatterers
surrounding the receiver antenna region. Here we see that modal
correlation at the receiver is independent of the mode selected from
transmitter region.

Similarly, we can write the correlation between $m$-th and
$m^{\prime}\!$-th modes at the transmitter region due to the
$\ell$-th mode at the receiver region as
\begin{align}\label{eqn:tx_modal_corr_coef}
\gamma_{m,m^{\prime}}&={\int_{\mathbb{S}^1}}{\powd}_{Tx}(\aod)e^{i(m-m^{\prime})\aod}d\aod,
\quad\forall{\,\,\ell},
\end{align}
where ${\powd}_{Tx}(\aod)={\int}G(\aod,\aoa){d\aoa}$ is the
normalized azimuth power distribution at the transmitter region. As
for the receiver modal correlation, we can observe that modal
correlation at the transmitter is independent of the mode selected
from receiver region. Note that, azimuth power distributions
${\powd}_{Rx}(\aoa)$ and ${\powd}_{Tx}(\aod)$ can be modeled using
all common power distributions such as uniform-limited
\cite{Salz-1994-uniform-dist}, Gaussian
\cite{Kalkan-1997-Gaus-dist}, Laplacian
\cite{Pedersen-2000-Laplace-dist}, $\cos^{2p}\phi$ distribution
\cite{Salz-1994-uniform-dist}, etc.

Denoting the $p$-th column of scattering matrix $\vH_S$ as
$\vH_{S,p}$, the $(2\RxM+1)\times(2\RxM+1)$ receiver modal
correlation matrix can be defined as
\begin{align}\nonumber
\vM_R &\triangleq\E{\vH_{S,p}\mct{\vH}_{S,p}},
\end{align}
where $(\ell,\ell^{\prime})$-th element of $\vM_R$ is given by
(\ref{eqn:rx_modal_corr_coef}) above. Similarly, the transmitter
modal correlation matrix can be defined as
\begin{align}\nonumber
\vM_T &\triangleq\E{\mct{\vH}_{S,q}\vH_{S,q}},
\end{align}
where $\vH_{S,q}$ is the $q$-th row of $\vH_S$. $(m,m^{\prime})$-th
element of $\vM_T$ is given by (\ref{eqn:tx_modal_corr_coef}) and
$\vM_T$ is a $(2\TxM+1)\times(2\TxM+1)$ matrix.

\subsubsection{Kronecker Model as a Special Case}
The correlation between two distinct modal pairs can be written as
the product of corresponding modal correlation at the transmitter
and the modal correlation at the receiver, i.e.,
\begin{align}\label{eqn:prod_kron_cor}
\gamma_{m,m^{\prime}}^{\ell,\ell^{\prime}}&=\gamma^{\ell,\ell^{\prime}}
\gamma_{m,m^{\prime}}.
\end{align}
Facilitated by \eqref{eqn:prod_kron_cor}, we write the covariance
matrix of the scattering channel $\vH_S$ as the Kronecker product
between the receiver modal correlation matrix and the transmitter
modal correlation matrix,
\begin{align}\label{eqn:prod_kron_matrix1}
\vR_S=\E{\mct{\vh}_S\vh_S}=\vM_R\otimes\vM_T.
\end{align}
Note that \eqref{eqn:prod_kron_cor} holds only for class of
scattering environments where the power spectral density of modal
correlation function satisfies
\cite{pollock_kronecker,Kermoal-2002-MIMO}
\begin{align}\label{eqn:power_product}
G(\aod,\aoa) = {\powd}_{Tx}(\aod){\powd}_{Rx}(\aoa).
\end{align}
Note that, \eqref{eqn:power_product} is the necessary condition in
which a channel must satisfy in order for \eqref{eqn:prod_kron_matrix1} to hold .\\

Assuming $\vR_S$ is a positive definite matrix, a channel
realization of the scattering channel $\vH_S$ can be generated by
\begin{align}\label{eqn:channel_simulation}
\mvec{(\vH_S)} &= \vR_S^{1/2}\mvec{(\vW_{S})},
\end{align}
where $\vR_S^{1/2}$ is the positive definite matrix square root
\cite{matrix_computations} of $\vR_S$ and $\vW_S$ is a
$(2\RxM+1)\times(2\TxM+1)$ matrix which has zero-mean independent
and identically distributed complex Gaussian random entries with
unit variance. Furthermore, using \eqref{eqn:prod_kron_matrix1}, the
full correlation matrix of the MIMO channel $\vH$, given by
\eqref{eqn:channel_decompo}, can be written as
\begin{align}\label{eqn:full_channel_cov}
\vR&=
\left(\con{\vJ}_R\vM_R\mt{\vJ}_R\right)\otimes\left(\vJ_T\vM_T\mct{\vJ}_T\right).
\end{align}

For simplicity, here we only consider the modal correlation at the
transmitter region and assume the effective communication modes
available at the receiver region are uncorrelated, i.e.
$\vM_{Rx}=\vI_{2\RxM+1}$. It was shown in
\cite{pollock_2003_space_mimo} that all azimuth power distribution
models give very similar correlation values for a given angular
spread, especially for small antenna separations. Therefore, without
loss of generality, we restrict our investigation only to the
uniform-limited azimuth power distribution, which is defined as
follows:\\

\textbf{\textit{Uniform-limited Azimuth Power Distribution}:} When
the energy is departing uniformly to a restricted range of azimuth
angles $\pm\bigtriangleup$ around a mean angle of departure (AOD)
$\maod\in[-\pi,\pi)$, we have the uniform-limited azimuth power
distribution \cite{Salz-1994-uniform-dist}
\begin{align}\nonumber
\mathcal{P}(\aod) &=
\frac{1}{2\bigtriangleup},\quad|\aod-\maod|\leq{\bigtriangleup},
\end{align}
where $\bigtriangleup$ represents the non-isotropic parameter of the
azimuth power distribution, which is related to the standard
deviation of the distribution (angular spread
$\sigma_t=\bigtriangleup/\sqrt{3}$). For the above APD, the
$(m,m')$-th entry of $\vM_T$ is given by
\begin{align}\nonumber
\left\{\vM_T\right\}_{m,m'}&=\mathrm{sinc}((m-m')\bigtriangleup)e^{i(m-m')\maod}.
\end{align}

\begin{figure}
\centering
  \includegraphics[width=\pictwidth]{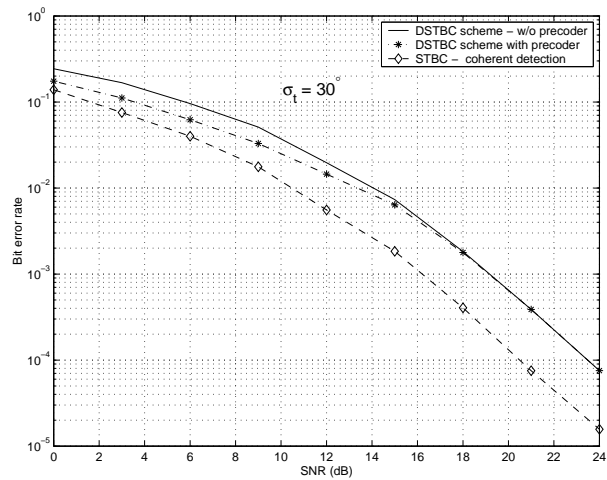}
  \caption{Precoder performance in non-isotropic
  scattering environments, $\sigma_t = 30^\circ$ mean AOD $\maod=0^\circ$ for a uniform-limited
  azimuth power distribution at the transmitter. $2\times{2}$ MIMO system. Transmit antenna
  separation $0.1\lambda$ and receive antenna separation $\lambda$: rate-$1$ differential STBC.
  }\label{fig:2tx_2rx_diff_angular_sp_30}
\end{figure}

\begin{figure}
\centering
  \includegraphics[width=\pictwidth]{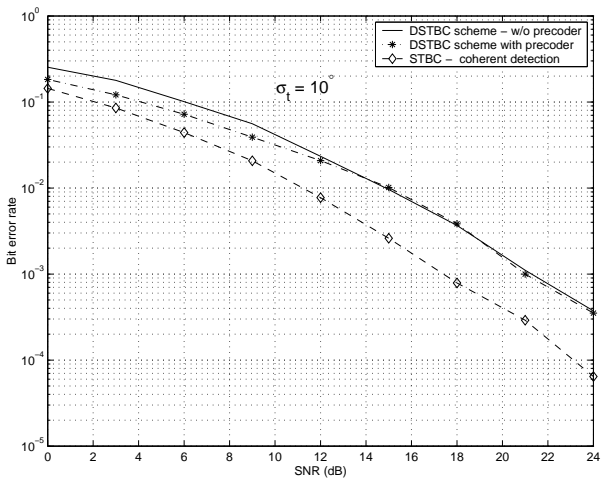}
  \caption{Precoder performance in non-isotropic
  scattering environments, $\sigma_t = 10^\circ$ mean AOD $\maod=0^\circ$ for a uniform-limited
  azimuth power distribution at the transmitter. $2\times{2}$ MIMO system. Transmit antenna
  separation $0.1\lambda$ and receive antenna separation $\lambda$: rate-$1$ differential STBC.
  }\label{fig:2tx_2rx_diff_angular_sp_10}
\end{figure}

Figures \ref{fig:2tx_2rx_diff_angular_sp_30} and
\ref{fig:2tx_2rx_diff_angular_sp_10} show the BER performance of
rate-1 differential STBC code with two transmit antennas for the
spatial arrangement considered in Section
\ref{sec:sim_results_MIMO_diff} for transmitter angular spreads
$\sigma_t=30^\circ$ and $10^\circ$ about the mean AOD
$\maod=0^\circ$. The channel is modeled using
\eqref{eqn:channel_decompo} and \eqref{eqn:channel_simulation}.

From Figures \ref{fig:2tx_2rx_diff_angular_sp_30} and
\ref{fig:2tx_2rx_diff_angular_sp_10} it is observed that in the
presence of non-isotropic scattering at the transmitter, proposed
precoding scheme provides significant BER improvements at low SNRs.
To further improve the performance, following Section
\ref{sec:problem_setup_diff_stbc}, a precoding scheme can be easily
derived by including the non-isotropic scattering parameters
(angular spreads and mean AOA/AOD) at both ends of the MIMO channel.
Unlike in the fixed precoding scheme, modified scheme will require
the receiver to estimate and feedback scattering distribution
parameters to the transmitter whenever there is a change in these
parameters.

\section{Performance in other Channel Models}\label{sec:performance_other_channels}
Simulation results presented in previous sections used the channel
model $\vH=\vJ_R\vH_S\mct{\vJ}_T$, which is derived based on plane
wave propagation theory, to simulate the underlying channels between
transmit and receive antennas. In this section we analyze the
performance of precoding schemes (coherent and differential) derived
in this paper applied on other \textit{statistical channel models}
proposed in the literature. In particular we are interested on
channel models that are consistent with wave propagation. MISO and
MIMO channel models proposed by Chen et al. \cite{chen_channel} and
Abdi et al. \cite{abdi_channel}, respectively are two such example
channel models. Sections \ref{sec:chen_channel_performance} and
\ref{sec:abdi_channel_performance} provide simulation results of
coherent STBC applied on Chen's MISO channel model and differential
STBC applied on Abdi's MIMO channel model, respectively. In
following simulations, precoders are derived using $\vJ_T$ and
$\vJ_R$ for given antenna configurations and the underlying channel
$\vH$ is simulated using Chen et al. and Abdi et al. channel models.

\begin{figure}
\centering
  \includegraphics[width=\pictwidth]{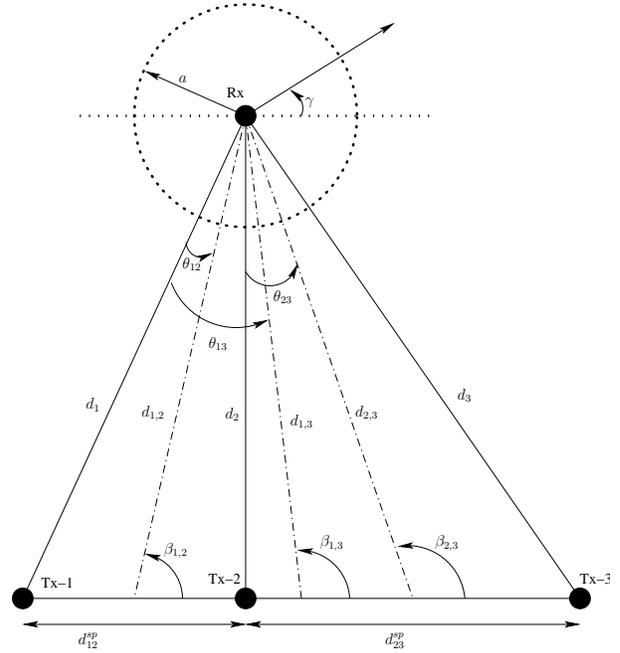}
  \caption{Scattering channel model proposed by Chen \textit{et al.} for three transmit and one receive
  antennas.}\label{fig:chen_channel}
\end{figure}

\subsection{Chen et al.'s MISO Channel
Model}\label{sec:chen_channel_performance} Fig.
\ref{fig:chen_channel} depicts the MISO channel model proposed by
Chen et al., where the space-time cross correlation between two
antenna elements at the transmitter is given by
\begin{small}
\begin{align}
\left[\vR(\tau)\right]_{m,n} &=
\exp\left[j\frac{2\pi}{\lambda}(d_m-d_n)\right]\times\label{eqn:chen_channel_corralation}\\
&\!\!\!\!\!\!\!\!\!\!\!\!\!\!\!\!J_0\left[2\pi\sqrt{\left(f_D\tau\cos\gamma+
\frac{z_{mn}^c}{\lambda}\right)^2+\left(f_D\tau\sin\gamma-\frac{z_{mn}^s}{\lambda}\right)^2}
\right]\nonumber
\end{align}
\end{small}
with
\begin{align}
z_{mn}^c &= \frac{2a}{d_m+d_n}\left[d_{mn}^{sp}-(d_m-d_n)\cos
\alpha_{mn}\cos \beta_{mn}\right],\nonumber\\
z_{mn}^s &= \frac{2a}{d_m+d_n}(d_m-d_n)\cos \alpha_{mn}\sin
\beta_{mn},\nonumber
\end{align}
$a$ is the scatterer ring radius, $\gamma$ is the moving direction
of the receiver with respect to the end-fire of the antenna, $f_D$
is the Doppler spread and $d_{mn}$ is the receiver distance to the
center of the transmit antenna pair $m,n$. All other geometric
parameters are defined as in Fig. \ref{fig:chen_channel}.

\begin{figure}[h]
\centering
\includegraphics[width=\pictwidth]{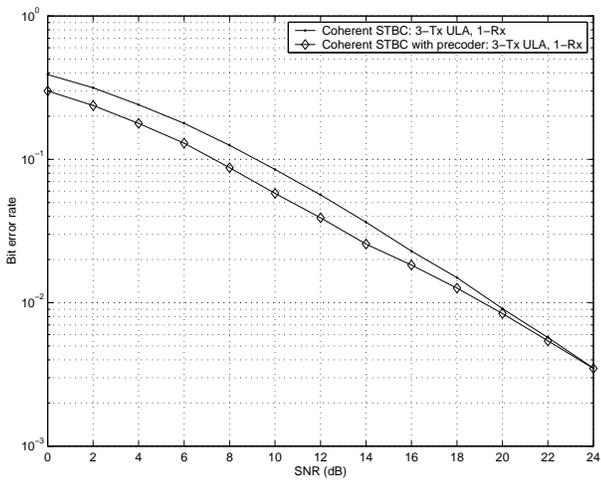}
\caption{Spatial precoder performance with three transmit and one
receive antennas for $0.2\lambda$ minimum separation between two
adjacent transmit antennas placed in a uniform linear array, using
Chen et al's channel model: rate-$3/4$ coherent STBC.}
\label{fig:3tx_1rx_coherent_chen}
\end{figure}

Fig. \ref{fig:3tx_1rx_coherent_chen} shows the performance of
spatial precoder derived in Section
\ref{sec:optimum_precoder_coherent_stbc} for rate-$3/4$ coherent
STBC with three transmit antennas placed in a ULA configuration. In
this simulation, we assume the time-varying channels are undergone
Rayleigh fading at the fading rate $f_DT=0.001$, where $T$ is the
codeword period. We set parameters $a=30\lambda$,
$d_{12}^{sp}=d_{23}^{sp}=0.2\lambda$, $d_{12}=1000\lambda$,
$\gamma=20^\circ$ and $\beta_{1,2}=60^\circ$. All other geometric
parameters of the model in Fig. \ref{fig:chen_channel} can be easily
determined from these parameters by using simple trigonometry. In
this simulation, a realization of the underlying space-time MIMO
channel is generated using \eqref{eqn:channel_simulation} and
\eqref{eqn:chen_channel_corralation}. From Fig.
\ref{fig:3tx_1rx_coherent_chen} we observed that proposed spatial
precoding scheme gives significant performance improvements for
time-varying channels. For example, at 0.05 BER, performance of the
spatially precoded system is $1$dB better than that of the
non-precoded system.

\begin{figure}[h]
\centering
  \includegraphics[width=\pictwidth]{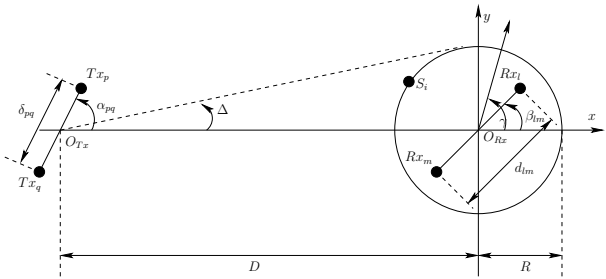}
  \caption{Scattering channel model proposed by Abdi \textit{et al.} for two transmit and two receive
  antennas.}\label{fig:abdi_channel}
\end{figure}

\subsection{Abdi et al.'s MIMO Channel Model}\label{sec:abdi_channel_performance}
In this model, space-time cross correlation between two distinct
antenna element pairs at the receiver and transmitter is given by
\begin{align}
\left[\vR(\tau)\right]_{l{p},mq} = &\frac{\exp[jc_{pq}\cos
(\alpha_{pq})]}{I_0(\kappa)}\times\nonumber\\
&I_0\left(\left\{\kappa^2-a^2-b_{lm}^2-c_{pq}^2\Delta^2\sin^2(\alpha_{pq})\right.\right.\nonumber\\
&\quad+2ab_{lm}\cos(\beta_{lm}-\gamma)+2c_{pq}\Delta\sin(\alpha_{pq})\nonumber\\
&\quad\times\left[a\sin(\gamma)-b_{lm}\sin(\beta_{lm})\right]\nonumber\\
&\quad-j2\kappa\left[a\cos(\mu-\gamma)-b_{lm}\cos(\mu-\beta_{lm})\right.\nonumber\\
&\quad-c_{pq}\Delta\sin(\alpha_{pq})\sin(\mu)\left.\left.\left.\right])\right\}^{1/2}\right),
\label{eqn:abdi_channel_corralation}
\end{align}
where $a=2{\pi}f_D\tau$, $b_{lm}=2{\pi}d_{lm}/\lambda$,
$c_{pq}=2{\pi}\delta_{pq}/\lambda$; $f_D$ is the Doppler shift;
$\mu$ is the mean angle of arrival at the receiver; $\kappa$
controls the spread of the AOA; and $\gamma$ is the direction of
motion of the receiver. Other geometric parameters are defined in
Fig. \ref{fig:abdi_channel}. Note that this model also captures the
non-isotropic scattering at the transmitter via $\Delta$ and the
model is valid only for small $\Delta$ \cite{abdi_channel}.

\begin{figure}
\centering
  \includegraphics[width=\pictwidth]{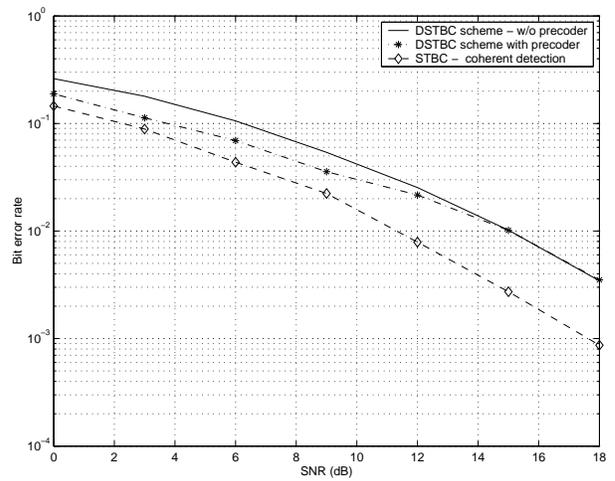}
  \caption{Spatial precoder performance with two transmit and two receive
  antennas using Abdi et al's channel model: rate-$1$ differential STBC.
  }\label{fig:2tx_2rx_diff_abdi}
\end{figure}

Fig. \ref{fig:2tx_2rx_diff_abdi} shows the performance of spatial
precoder derived in Section \ref{sec:optimum_precoder_diff_stbc} for
rate-$1$ differential STBC with two transmit and two receive
antennas for a stationary receiver (i.e. $f_D=0$). In this
simulation we set $\delta_{12}=0.1\lambda$, $d_{12}=\lambda$ and
$\alpha_{12}=\beta_{12}=0^\circ$. We assume the scattering
environment surrounding the receiver antenna array is rich, i.e.,
$\kappa=0$ and the non-isotropic factor $\Delta$ at the transmitter
is $10^\circ$. We assume the scattering channel satisfies the power
distribution condition \eqref{eqn:power_product}. A realization of
the underlying MIMO channel is generated using
\eqref{eqn:channel_simulation} and
\eqref{eqn:abdi_channel_corralation}. It is observed that our
precoding scheme based on antenna configuration details give
promising improvements for low SNR when the underlying channel is
modeled using Abdi's channel model. Therefore, using the previous
results from Chen's channel model and the current results, we can
come to the conclusion that our fixed spatial precoding scheme can
be applied to any general wireless communication system.

\section{Concluding Remarks}\label{sec:conclusion}
In this paper, by exploiting the spatial dimension of a MIMO channel
we have proposed spatial precoding schemes for coherent and
differential space-time block coded systems. Precoders are derived
by minimizing certain upper bounds for the PEP subject to a transmit
power constraint and assuming an isotropic scattering environment
surrounding the transmit and receive antenna arrays. The proposed
precoders are designed based on previously unutilized fixed and
known parameters of MIMO channels, the antenna spacing and antenna
placement details. Therefore, with these schemes the transmitter
does not require any feedback of channel state information from the
receiver, which is an added advantage over the other precoding
schemes found in the literature. Since the precoder is fixed for
fixed antenna configurations, proposed precoding schemes can be
applied in non-stationary scattering channels as well as stationary
scattering channels.

We showed that proposed precoding schemes reduce the detrimental
effects of non-ideal antenna placement and improve the performance
of space-time coded MIMO systems. Precoders achieve these
performance improvements by virtually arranging antennas into an
optimal configuration as such the spatial correlation between all
antenna elements is minimum. For 1-D arrays (ULA), we observed that
precoder gives scope for improvement at high SNRs, but for 2-D
arrays (UCA), improvements are only seen at low SNRs.

Although the proposed precoders are derived for isotropic scattering
environments, we observed that these precoders give significant
performance improvements in non-isotropic scattering environments.
Based on the performance improvements we observed, we believe that
proposed schemes can be applied on uplink transmission of a mobile
communication system as the proposed schemes can effectively reduce
the effects due to insufficient antenna spacing and antenna
placement at the mobile unit.


\appendices

\section{Proof of PEP Upper bound: Coherent
Receiver}\label{app:pep_bound_derivation_coherent} The conditional
average pairwise error probability
$\mathrm{P}(\vS_i\rightarrow\vS_j)$, defined as the probability that
the receiver erroneously decides in favor of $\vS_j$ when $\vS_i$
was actually transmitted for a given channel, is upper bounded by
the \textit{Chernoff bound} \cite{tarokh_1998_first}
\begin{align}\label{eqn:chernof_bound}
\mathrm{P}(\vS_i\rightarrow\vS_j|\vh)\:&{\leq}\:\mathrm{exp}\left(-\frac{\snr}{4}
d_h^2(\vS_i,\vS_j)\right),
\end{align}
where
$d_h^2(\vS_i,\vS_j)=\vh[\vI_{\RxA}\otimes\vS_{\Delta}]\mct{\vh}$,
$\vS_{\Delta}=\vF_d(\vS_i-\vS_j)\mct{(\vS_i-\vS_j)}\mct{\vF}_d$,
$\vh=(\mvec{(\vH^T)})^T$ a row vector and $\snr =
{E_s}/{\sigma_n^2}$ is the average SNR at each receiver antenna.  To
compute the average PEP, we average (\ref{eqn:chernof_bound}) over
the joint distribution of $\vh$. Assume $\vh$ is a proper
complex\footnote{To be proper complex, the mean of both the real and
imaginary parts of $\vH_S$ must be zero and also the
cross-correlation between real and imaginary parts of $\vH_S$ must
be zero.} $\TxA\RxA$-dimensional Gaussian random vector with mean
$\ve{0}$ and covariance matrix $\vR_{\vH}=\E{\mct{\vh}\vh}$, then
the pdf of $\vh$ is given by \cite{goodman_63}
\begin{align}\nonumber
p(\vh)&=
\frac{1}{\pi^{\TxA\RxA}\left|\vR_{\vH}\right|}\exp\{{-\vh\vR_{\vH}^{-1}\mct{\vh}}\},
\end{align}
provided that $\vR_{\vH}$ is non-singular. Then the average PEP is
bounded as follows
\begin{align}\label{eqn:gaussian_one}
\mathrm{P}(\vS_i\rightarrow\vS_j)\:&{\leq}
\frac{1}{\pi^{\TxA\RxA}\left|\vR_{\vH}\right|}\int\exp\{{-\vh\vR_0^{-1}\mct{\vh}}\}d\vh
\end{align}
where
$\vR_0^{-1}=(\frac{\snr}{4}\vI_{\RxA}\otimes\vS_{\Delta}+\vR_{\vH}^{-1})$.
Assume $\vR_{\vH}$ is non-singular (positive definite), therefore
the inverse $\vR_{\vH}^{-1}$ is positive definite, since the inverse
matrix of a positive definite matrix is also positive definite
\cite[page 142]{matrix_computations}. Also note that $\vS_{\Delta}$
is Hermitian and it has positive eigenvalues (through code
construction, e.g. \cite{tarokh_1998_first}), therefore
$\vS_{\Delta}$ is positive definite, hence
$\vI_{\RxA}\otimes\vS_{\Delta}$ is also positive definite. Therefore
$\vR_0^{-1}$ is positive definite and hence $\vR_0$ is non-singular.
Using the normalization property of Gaussian pdf
\begin{align}
\frac{1}{\pi^{\TxA\RxA}\left|\vR_0\right|}\int{\exp\{{-\vh\vR_0^{-1}\mct{\vh}}\}d\vh}
&=1,\nonumber
\end{align}
we can simplify (\ref{eqn:gaussian_one}) to
\begin{align}
\mathrm{P}(\vS_i\rightarrow\vS_j)\:&{\leq}\frac{\left|\vR_0\right|}{\left|\vR_{\vH}\right|}
\:=\frac{1}{\left|\vR_0^{-1}\vR_{\vH}\right|},\nonumber
\end{align}
or equivalently
\begin{align}\nonumber
\mathrm{P}(\vS_i\rightarrow\vS_j)\:&{\leq}\frac{1}{\left|\vI_{\TxA\RxA}+\frac{\snr}{4}
\vR_{\vH}[\vI_{\RxA}\otimes\vS_{\Delta}]\right|}.
\end{align}\\

\section{Proof of Generalized Water-filling Solution for $\RxA=2$ Receiver Antennas}
\label{app:two_rx}
Let $\RxA=2$ in \eqref{eqn:kkt_final_b_coherent}, then we obtain the
second-order polynomial
$r_1r_2{\upsilon_c}t_i^2q_i^2+({\upsilon_c}t_i(r_1+r_2)-2r_1r_2t_i^2)q_i+
({\upsilon_c}-r_1t_i-r_2t_i)$
in $q$ which has roots $q_{i,1}=A+\sqrt{K}$ and
$q_{i,2}=A-\sqrt{K}$, where $A$ and $K$ are given by
\eqref{eqn:A_K_coherent}. Then the product
$q_{i,1}q_{i,2}=({\upsilon_c}-r_1t_i-r_2t_i)/r_1r_2{\upsilon_c}t_i^2$.\\

\textbf{Case 1:} $q_{i,1}q_{i,2}>0$ $\Rightarrow$
${\upsilon_c}>t_i(r_1+r_2)$. In this case, both roots are either
positive or negative. Let ${\upsilon_c}={\alpha}t_i(r_1+r_2)$, where
$\alpha>1$. Then $A=-t_i^2\alpha[(r_1+r_2)^2-2r_1r_2/\alpha]<0$ for
all $\alpha>1$. Since $K>0$, $q_{i,2}<0$, thus $q_{i,1}$ must also
be negative to hold ${\upsilon_c}>t_i(r_1+r_2)$. Therefore, when
${\upsilon_c}>t_i(r_1+r_2)$, the optimum $q_i$ is zero
to hold the inequality constraints of \eqref{eqn:opt_prob_coherent1}.\\

\textbf{Case 2:} $q_{i,1}q_{i,2}<0$ $\Rightarrow$
${\upsilon_c}<t_i(r_1+r_2)$. In this case, we always have one
positive root and one negative root. Assume $q_{i,1}>0$ and
$q_{i,2}<0$ and let ${\upsilon_c}={\alpha}t_i(r_1+r_2)$, where
$0<\alpha<1$. For $q_{i,1}$ to positive, we need to prove that
$\sqrt{K}>t_i^2\alpha[(r_1+r_2)^2-2r_1r_2/\alpha]$ for $0<\alpha<1$.
Instead, we show that
\begin{align}
\sqrt{K}<t_i^2\alpha[(r_1+r_2)^2-2r_1r_2/\alpha]\label{eqn:_proof_1_1},
\end{align}
only when $\alpha>1$. Note that, since $K>0$, \eqref{eqn:_proof_1_1}
can be squared without affecting to the inequality sign. Therefore
squaring \eqref{eqn:_proof_1_1} and further simplification to it
yields $\alpha>1$. This proves that $q_{i,1}>0$ and $q_{i,2}<0$ when
${\upsilon_c}<t_i(r_1+r_2)$ and the optimum solution to
\eqref{eqn:opt_prob_coherent1} is given by $q_{i,1}$.

\section{Proof of Generalized Water-filling Solution for $\RxA=3$ Receiver Antennas}
\label{app:three_rx}
Let $\RxA=3$ in \eqref{eqn:kkt_final_b_coherent}, then we obtain the
third-order polynomial $a_3q_i^3+a_2q_i^2+a_1q_i+a_0$ in $q_i$ which
has roots \cite{math_handbook_1968}
\begin{align}
q_{i,1}&=-\frac{a_2}{3} +(S+T),\nonumber\\
q_{i,2}&=-\frac{a_2}{3}
-\frac{1}{2}(S+T)+\frac{\imath\sqrt{3}}{2}(S-T),\nonumber\\
q_{i,3}&=-\frac{a_2}{3}
-\frac{1}{2}(S+T)-\frac{\imath\sqrt{3}}{2}(S-T),\nonumber
\end{align}
where $S\pm{T} = \left[R+\sqrt{Q^3+R^2}\right]^{\frac{1}{3}}\pm
\left[R-\sqrt{Q^3+R^2}\right]^{\frac{1}{3}}$ and all other variables
are as defined in Section \ref{sec:miso_channel_sol_ntx3_coherent},
then the product
$q_{i,1}q_{i,2}q_{i,3}=(r_1t_i+r_2t_i+r_3t_i-{\upsilon_c})/r_1r_2r_3{\upsilon_c}t_i^3.$\\

\textbf{Case 1:} $q_{i,1}q_{i,2}q_{i,3}<0$ $\Rightarrow$
${\upsilon_c}>t_i(r_1+r_2+r_3)$. Let
${\upsilon_c}={\alpha}t_i(r_1+r_2+r_3)$, where $\alpha>1$. For
$\alpha>1$, it can be shown that $(Q^3+R^2)>0$, hence $q_{i,1}<0$
and $q_{i,2},q_{i,3}\in\mathbb{C}$. Therefore,
when ${\upsilon_c}>t_i(r_1+r_2+r_3)$, the optimum $q_i$ is zero.\\

\textbf{Case 2:} $q_{i,1}q_{i,2}q_{i,3}>0$ $\Rightarrow$
${\upsilon_c}<t_i(r_1+r_2+r_3)$. Let
${\upsilon_c}={\alpha}t_i(r_1+r_2+r_3)$, where $0<\alpha<1$. For
$0<\alpha<1$, it can be shown that $(Q^3+R^2)<0$ and
$R^{\frac{1}{3}}>\frac{a_2}{6}$, hence we get two negative roots
$q_{i,2},q_{i,3}<0$ and one positive root $q_{i,1}>0$ as the roots
of cubic polynomial. Therefore, when
${\upsilon_c}<t_i(r_1+r_2+r_3)$, the optimum solution to
\eqref{eqn:opt_prob_coherent1} is given by $q_{i,1}$.\\

\section{Proof of the conditional mean and the conditional variance of
$u=2\mathrm{Re}\{\vw(k)\mct{\vDel}_{i,j}\mct{\vy}(k-1)\}$}
\label{app:mean_and_var_of_u}
\subsection{Proof of Conditional Mean}
Mean of $u$ condition on the received signal $\vy(k-1)$ can be
written as
\begin{align}\nonumber
\bar{m}_{u{\mid}\vy(k-1)}
&=\E{2\mathrm{Re}\left\{\vw(k)\mct{\vDel}_{i,j}\mct{\vy}(k-1)\right\}\mid\vy(k-1)},\nonumber\\
&=2\mathrm{Re}\left\{\E{\vw(k)\mid\vy(k-1)}\mct{\vDel}_{i,j}\mct{\vy}(k-1)\right\}.\label{eqn:cond_mean_1}
\end{align}
Substituting $\vw(k) = \vn(k)-\vn(k-1)\vSm_{i}$ and noting
$\E{\vn(k)\mid\vy(k-1)}=0$, \eqref{eqn:cond_mean_1} can be
simplified to
\begin{align}\nonumber
\bar{m}_{u{\mid}\vy(k-1)}
&=-2\mathrm{Re}\left\{\bar{m}_{\vn(k-1)\mid\vy(k-1)}\vSm_{i}\mct{\vDel}_{i,j}\mct{\vy}(k-1)\right\},
\nonumber\\
&=2\mathrm{Re}\left\{\bar{m}_{\vn(k-1)\mid\vy(k-1)}(\vI-\vSm_{i}\mct{\vSm}_j)\mct{\vy}(k-1)\right\},
\label{eqn:cond_mean_2}
\end{align}
where $\bar{m}_{\vn(k-1)\mid\vy(k-1)}=\E{\vn(k-1)\mid\vy(k-1)}$.
Using the minimum mean square error estimator results given in
\cite[Section 2.3]{anderson_moore}, we obtain
\begin{align}
\bar{m}_{\vn(k-1)\mid\vy(k-1)} = \,& \E{\vn(k-1)} +
\left[\vy(k-1)-\E{\vy(k-1)}\right]\nonumber \\
&\times\Sigma_{\vy(k-1),\vy(k-1)}^{-1}\Sigma_{\vy(k-1),\vn(k-1)},\nonumber
\end{align}
where
\begin{align}
\Sigma_{\vy(k-1),\vy(k-1)}&= \E{\mct{\vy}(k-1)\vy(k-1)},\label{eqn:sig_yk-1_yk-1}\\
&=E_s\mct{\vXm(k-1)}\vR_{\vH}\vXm(k-1)+\sigma_n^2\vI_{\TxA\RxA},\nonumber
\end{align}
and
\begin{align}
\Sigma_{\vy(k-1),\vn(k-1)}&=\E{\mct{\vy}(k-1)\vn(k-1)},\nonumber\\
&=\sigma_n^2\vI_{\TxA\RxA}.\label{eqn:sig_yk-1_nk-1}
\end{align}

Since $\E{\vn(k-1)}=0$ and $\E{\vy(k-1)}=0$, we have
\begin{align}
\bar{m}_{\vn(k-1)\mid\vy(k-1)} = &\,\,
\sigma_n^2\vy(k-1)\label{eqn:noise_con_yk}\\
&\!\!\!\times\left(E_s\mct{\vXm(k-1)}\vR_{\vH}\vXm(k-1)+\sigma_n^2\vI\right)^{-1}\!\!\!.\nonumber
\end{align}
Substituting \eqref{eqn:noise_con_yk} for
$\bar{m}_{\vn(k-1)\mid\vy(k-1)}$ in \eqref{eqn:cond_mean_2} gives
the conditional mean $\bar{m}_{u{\mid}\vy(k-1)}$.

\subsection{Proof of Conditional Variance}
Variance of $u$ condition on the received signal $\vy(k-1)$ can be
written as
\begin{align}
\sigma_{u{\mid}\vy(k-1)}^2
&=\E{{\parallel}u-\bar{m}_{u{\mid}\vy(k-1)}\parallel^2\mid\vy(k-1)}\label{eqn:cond_var_1}\\
&=\E{\mct{(u-\bar{m}_{u{\mid}\vy(k-1)})}(u-\bar{m}_{u{\mid}\vy(k-1)})\mid\vy(k-1)}.\nonumber
\end{align}
After some straight forward manipulations we can show
\begin{align}\label{eqn:cond_var_2}
{u}-\bar{m}_{u{\mid}\vy(k-1)} &=
2\mathrm{Re}\left\{\left(\vn(k)-\left[\vn(k-1)-\bar{m}_{\vn(k-1)\mid\vy(k-1)}\right]
\right.\right.\nonumber\\
&\times\left.\left.\vSm_{i}\right)\mct{\vDel}_{i,j}\mct{\vy}(k-1)\right\}.
\end{align}

Substituting \eqref{eqn:cond_var_2} for
${u}-\bar{m}_{u{\mid}\vy(k-1)}$ in \eqref{eqn:cond_var_1} gives
\eqref{eqn:cond_var_3}, shown at the top of the next page, where
\begin{figure*}[bt]
 \normalsize
\begin{subequations}\label{eqn:cond_var_3}
\begin{align}
\sigma_{u{\mid}\vy(k-1)}^2
=&\;\mathrm{E}\left\{\mct{\left[2\mathrm{Re}\left\{\left(\vn(k)-\left[\vn(k-1)-
\bar{m}_{\vn(k-1)\mid\vy(k-1)}\right]\vSm_{i}\right)\mct{\vDel}_{i,j}\mct{\vy}(k-1)\right\}
\right]}\right.\nonumber\\
&\times\left[2\mathrm{Re}\left.\left\{\left(\vn(k)-\left[\vn(k-1)-
\bar{m}_{\vn(k-1)\mid\vy(k-1)}\right]\vSm_{i}\right)\mct{\vDel}_{i,j}\mct{\vy}(k-1)\right\}\right]
\mid\vy(k-1)\right\},\label{eqn:cond_var_3a}\\
=&\;2\vy(k-1)\vDel_{i,j}\left[\Sigma_{\vn(k),\vn(k)}-\mct{\vSm}_{i}\Sigma_{\vn(k-1)\mid\vy(k-1)}
\vSm_{i}\right]\mct{\vDel}_{i,j}\mct{\vy}(k-1),\label{eqn:cond_var_3b}
\end{align}
\end{subequations}
\hrulefill
\end{figure*}
$\Sigma_{\vn(k),\vn(k)}=\E{\mct{\vn}(k)\vn(k)}=\sigma_n^2\vI$ and
\begin{align}
\Sigma_{\vn(k-1)\mid\vy(k-1)} =&\:\nonumber\\
&\!\!\!\!\!\!\!\!\E{{\parallel}\vn(k-1)-\bar{m}_{n(k-1){\mid}\vy(k-1)}\parallel^2
\mid\vy(k-1)}\nonumber
\end{align}
is the covariance of the noise vector $\vn(k-1)$ condition on
$\vy(k-1)$. Using the minimum mean square error estimator results
given in \cite{anderson_moore}, we can write
\begin{align}
\Sigma_{\vn(k-1)\mid\vy(k-1)} = &\;\;
\Sigma_{n(k-1),n(k-1)}\nonumber\\
&\!\!\!\!\!\!\!\!\!\!\!\!-\mct{\Sigma}_{\vy(k-1),\vn(k-1)}\Sigma_{\vy(k-1),\vy(k-1)}^{-1}
\Sigma_{\vy(k-1),\vn(k-1)},\nonumber\\
=
&\;\;\sigma_n^2\left[\vI-\sigma_n^2\Sigma_{\vy(k-1),\vy(k-1)}^{-1}\right]\label{eqn:cond_var_4}
\end{align}
Substituting \eqref{eqn:sig_yk-1_yk-1} for
$\Sigma_{\vy(k-1),\vy(k-1)}$ in \eqref{eqn:cond_var_4} and then the
result in \eqref{eqn:cond_var_3b} gives the conditional variance
$\sigma_{u{\mid}\vy(k-1)}^2$.

\section{Proof of PEP Upper bound: Non-coherent Receiver}
\label{app:pep_bound_derivation_noncoherent} At asymptotically high
SNRs, the PEP condition on the received signal $\vy(k-1)$ is given
by
\begin{align}\nonumber
\mathrm{P}(\vS_i\rightarrow\vS_j\mid\vy(k-1))
 &=\mathrm{Q}\left(\sqrt{\frac{d_{i,j}^2}{4\sigma_n^2}}\right).
\end{align}
Now using the Chernoff bound
\begin{align}\nonumber
\mathrm{Q}(x)\leq\frac{1}{2}\exp\left(\frac{-x^2}{2}\right),
\end{align}
the conditional PEP can be upper bounded by
\begin{align}\label{eqn:pep_cond_yk-1_2}
\mathrm{P}(\vS_i\rightarrow\vS_j\mid\vy(k-1))
 &\leq\frac{1}{2}\exp\left(\frac{-d_{i,j}^2}{8\sigma_n^2}\right).
\end{align}
To compute the average PEP, we average \eqref{eqn:pep_cond_yk-1_2}
over the joint distribution of $\vy(k-1)$. Assume $\vy(k-1)$ is a
proper complex Gaussian random vector that has mean
$\E{\vy(k-1)}=\ve{0}$ and covariance
\begin{align}\nonumber
\vR_{\vy(k-1)}\triangleq&\;\E{\mct{\vy}(k-1)\vy(k-1)},\\
=&\;E_s\mct{\vXm(k-1)}\vR_{\vH}\vXm(k-1)+\sigma_n^2\vI_{\TxA\RxA}
\end{align}

If $\vR_{\vy(k-1)}$ is non-singular, then the pdf of $\vy(k-1)$ is
given by
\begin{align}\nonumber
p(\vy(k-1))&=
\Omega_y\exp\left\{{-\vy(k-1)\vR_{\vy(k-1)}^{-1}\mct{\vy}(k-1)}\right\},
\end{align}
where $\Omega_y={\pi^{-\TxA\RxA}/\left|\vR_{\vy(k-1)}\right|}$.
Averaging \eqref{eqn:pep_cond_yk-1_2} over the pdf of $\vy(k-1)$, we
obtain
\begin{align}
\mathrm{P}(\vS_i\rightarrow\vS_j)
&\leq\nonumber\\
&\!\!\!\!\!\!\!\!\!\!\frac{\Omega_y}{2}\int
\exp\left\{{-\vy(k-1)\vR_{d}^{-1}\mct{\vy}(k-1)}\right\}\mathrm{d}\vy(k-1),\label{eqn:pep_cond_yk-1_3}
\end{align}
where
\begin{align}\nonumber
\vR_{d}^{-1}= \vR_{\vy(k-1)}^{-1} + \frac{1}{8\sigma_n^2}\vD_{i,j}.
\end{align}
Assume $\vR_{\vH}$ is non-singular (positive definite). It can be
shown that both $\vR_{\vy(k-1)}$ and $\vD_{i,j}$ are positive
definite. Therefore, $\vR_{d}$ is non-singular. Using the
normalization property of Gaussian pdf
\begin{small}
\begin{align}
\frac{1}{\pi^{\TxA\RxA}\left|\vR_d\right|}\int\exp\left\{{-\vy(k-1)\vR_{d}^{-1}
\mct{\vy}(k-1)}\right\}\mathrm{d}\vy(k-1)
&=1,\nonumber
\end{align}
\end{small}
we can simplify \eqref{eqn:pep_cond_yk-1_3} to
\begin{align}
\mathrm{P}(\vS_i\rightarrow\vS_j)\:&{\leq}\frac{\left|\vR_d\right|}{2\left|\vR_{\vy(k-1)}\right|}
\:=\frac{1}{2\left|\vR_d^{-1}\vR_{\vy(k-1)}\right|},\nonumber
\end{align}
or equivalently
\begin{align}
\mathrm{P}&(\vS_i\rightarrow\vS_j)
\leq&\nonumber\\
&\frac{1}{2}\frac{1}{\left|\vI+\frac{1}{8}
\left(\snr\mct{\vXm(k-1)}\vR_{\vH}\vXm(k-1)+\vI_{\TxA\RxA}\right)\vD_{i,j}\right|}.\nonumber
\end{align}

\end{document}